\documentclass[superscriptaddress,groupedaddress,nofootnoteinbib,11pt]{article}
\pdfoutput=1 % if your are submitting a pdflatex (i.e. if you have images in pdf, png or jpg format)
\usepackage{jheppub}

\usepackage{CJKutf8} % for Chinese characters
\usepackage{mathtools,slashed,mathrsfs}
\usepackage[caption=false]{subfig}
\usepackage{dcolumn}% Align table columns on decimal point
\usepackage{multirow}
\usepackage{tabularx}
\usepackage{booktabs}
\usepackage{bm}% bold math
\usepackage{amsmath}
\usepackage{comment}
\usepackage{setspace}
\usepackage[dvipsnames]{xcolor}
\usepackage[normalem]{ulem} % to strike through
\usepackage{enumerate}
\usepackage{siunitx}
\usepackage{float}
\usepackage{hyperref}
\usepackage{cancel}

\newcommand{\paren}[1]{\left( #1 \right)}

\newcommand{\Cite}[1]{ref.~\cite{#1}}
\newcommand{\fig}[1]{figure~\ref{#1}}
\newcommand{\eq}[1]{eq.~(\ref{#1})}
\newcommand{\eqs}[2]{eqs.~(\ref{#1}) and (\ref{#2})}
\newcommand{\Sec}[1]{section~\ref{#1}}

\newcommand{\app}[1]{appendix~\ref{#1}}

\title{\centering Grand Unification at the Cosmological Collider  \\ with Chemical Potential}

\date{\today}
\author[a,b]{Arushi Bodas,}
\author[c]{Edward Broadberry,}
\author[c]{Raman Sundrum}

\affiliation[a]{Enrico Fermi Institute, University of Chicago, Chicago, IL 60637, USA}
\affiliation[b]{Particle Theory Department, Fermilab, Batavia, Illinois 60510, USA}
\affiliation[c]{Maryland Center for Fundamental Physics, Department of Physics, University of Maryland, College Park, MD 20742, USA}

\emailAdd{arushib@uchicago.edu}
\emailAdd{edbroad@umd.edu}
\emailAdd{raman@umd.edu}

\abstract{We introduce a tree-level 
 chemical potential mechanism for spin-$1$ particles within cosmological collider physics, allowing them to be detected in primordial non-Gaussianities for masses above the inflationary Hubble scale. We apply this mechanism to orbifold grand unification and the massive unification partners of the standard model gauge bosons. Our mechanism requires at least a pair of massive vector fields which are singlets of the standard model, a condition which is satisfied in the classic ``trinification'' scenario. Assuming that the gauge hierarchy problem is solved by supersymmetry, 
 gauge coupling running points to unification partners at $\sim 10^{15}$ GeV. We show that, within high-scale inflation, chemical potential enhancement can lead to observably strong signals for trinification partners in future cosmological surveys.

}

\begin{document}
% \hspace{25em} {\small FERMILAB-PUB-24-0568-V}

\maketitle

\section{Introduction}

The emerging field of cosmological collider physics \cite{CCP1,CCP3,CCP6,CCP2,CCP4,arkanihamed2015cosmological,Lee_2016} shows how primordial correlators can probe particle physics at inflationary energy scales, which may be many orders of magnitude above that of any terrestrial collider. Current data places an upper bound on the Hubble scale during inflation \cite{Planck}
\begin{equation}
    H \lesssim 5 \times 10^{13} \textrm{ GeV},
    \label{eq:HubbleBound}
\end{equation}
informing us of the potential reach of the cosmological collider.
The core universal mechanism involves particle production due to spacetime expansion for all species with masses $m\sim O(H)$. If such heavy particles decay into inflatons their propagation will be imprinted on primordial non gaussianities (NGs), with potentially observable signals in CMB data (see \cite{PlanckCCP} for recent data analysis), large scale structure surveys \cite{LSS1,LSS2,LSS3,LSS4} (see \cite{CCPconstraints} for recent data analysis), and high-redshift 21-cm tomography \cite{21cm1,21cm2}.

The proto-typical signal for on-shell exchange of a heavy particle is given by \cite{CCP1,CCP2}
\begin{equation}    
    \label{eq:typicalsignal}
    \frac{\langle\mathcal{R}_{\vec{k}_1}\mathcal{R}_{\vec{k}_2}\mathcal{R}_{\vec{k}_3}\rangle'}{\langle\mathcal{R}_{\vec{k}_1}\mathcal{R}_{-\vec{k}_1}\rangle'\langle\mathcal{R}_{\vec{k}_3}\mathcal{R}_{-\vec{k}_3}\rangle'} \approx e^{-\pi\mu_0}\left(f(\mu_0)\left(\frac{k_1}{k_3}\right)^{-3/2 + i\mu_0}+c.c.\right),
\end{equation}
%The focus of this paper is the on-shell production of a spin-1 particle with mass $m$. Previous papers have found a typical signal of the form \cite{Kumar:2017ecc, Kumar:2018jxz, Lee_2016}
where $\mu_0 =\sqrt{m^2/H^2 -9/4}$, and $\mathcal{R}$ is the comoving curvature perturbation. 
The key observable feature is the non-analytic dependence on comoving momentum, $k$. However, inspection of \eq{eq:typicalsignal} shows the mass must be in a narrow window around $m \sim H$ due to an exponential suppression for large masses, and the non-analytic dependence being unobservable for small masses. 
Intuitively, the exponential suppression for large masses can be thought of as a Boltzmann suppression, where the exponential factor is the square root of a Boltzmann probability for particle excitation with respect to a Hawking temperature of dS-space $T_{\rm H} = H/2\pi$ \cite{Birrell:1982ix}.

Since observability requires a new particle's mass to lie in a narrow range, it is important to consider whether new particles are expected at this scale. 
A highly motivated class of models predicting new particles at around this energy are grand unified theories (GUTs) \cite{guts1,guts2} (see \cite{ParticleDataGroup:2022pth} for a review). 
The main piece of quantitative evidence for GUTs is the approximate unification of the SM gauge couplings if their RG flow is extrapolated to energies of $M_{\rm GUT}\sim 10^{14}$ GeV. 
However, the minimal GUT scenarios predict proton decay at a rate ruled out by experiments \cite{ProtDec1,ProtDec2}. 
Proton decay can be naturally avoided in higher-dimensional ``orbifold GUTs" where the GUT symmetry is broken to the standard model by boundary conditions on a $(3+1)$-dimensional boundary \cite{Kawamura:1999nj,Kawamura:2000ev,Hall1}. 
Ref. \cite{Kumar:2018jxz} investigated the cosmological collider signals of such models. 
It was found that the signals for the heavy gauge bosons were highly suppressed, so the largest contribution to NGs was a Planck suppressed signal from the Kaluza Kelin (KK) graviton. 
Remarkably the KK graviton had small but observable signals. 
However, this required an even narrower mass range for these spin-2 particles than in the spin-0 case of \eq{eq:typicalsignal}.

\begin{figure}[t]
\centering
\includegraphics[width=.75\textwidth]{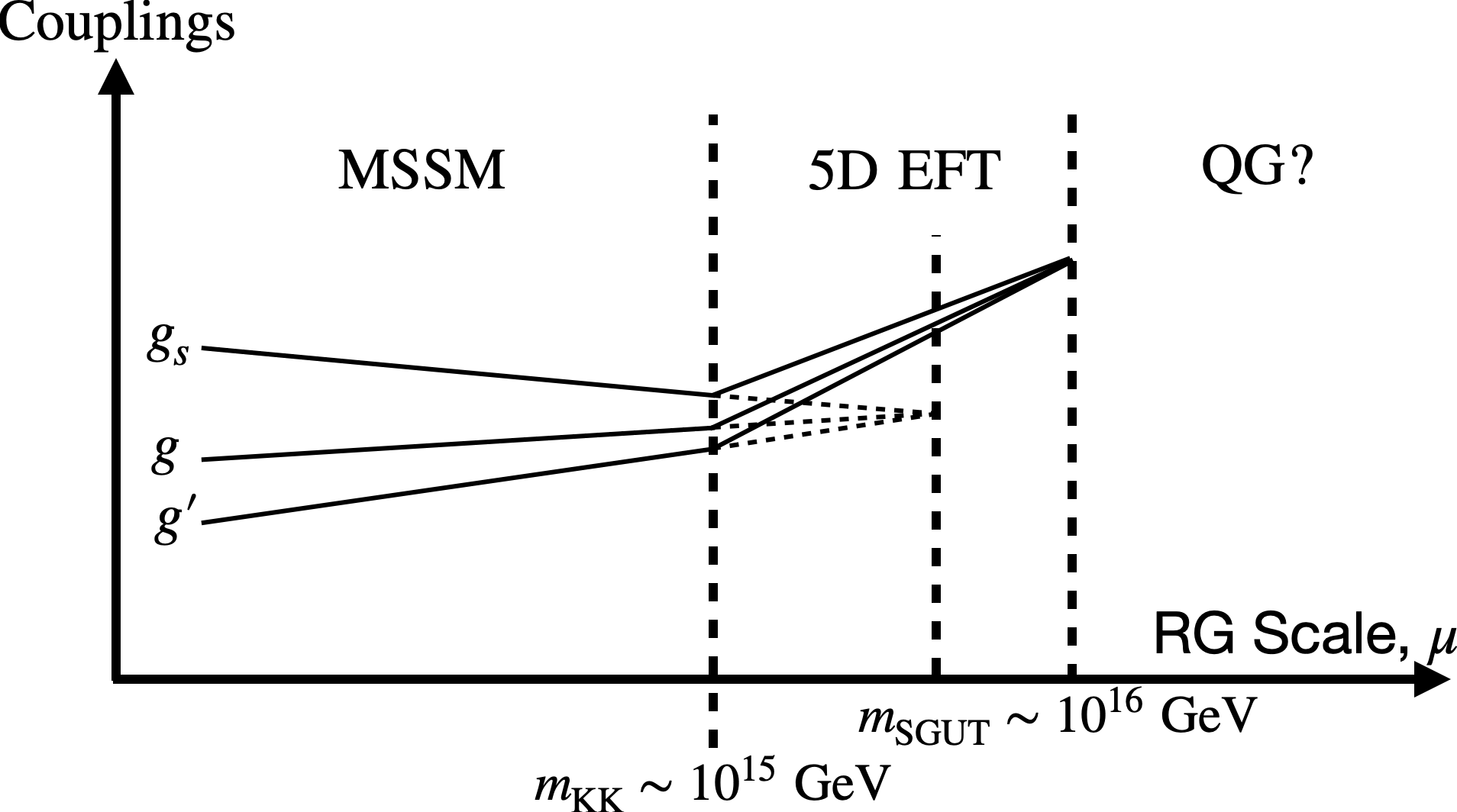}
\qquad
\caption{Adapted from \Cite{Hall1}. In  orbifold GUTs the gauge couplings need not exactly unify before the extra dimension becomes important. Above this scale the running is modified, and it has been speculated that they unify with the gravitational coupling into a strongly coupled string theory, or some other UV completion of quantum gravity.
\label{fig:Orbifold coupling unification}}
\end{figure}

Minimal GUTs have a few theoretical drawbacks. Firstly, the quantitative evidence in the form of the coupling unification is not so precise. 
Secondly, the value of the electroweak symmetry breaking (EWSB) scale $v \ll M_{\rm GUT}$ requires fine tuning, that is the gauge hierarchy problem. 
Both of these problems are solved in supersymmetric (SUSY) GUTs where the coupling unification is more precisely achieved, and a small EWSB scale is radiatively stable. 
However, bounds on proton decay require the colored GUT partners of the Higgs doublet to be at the GUT scale. 
This is called the doublet-triplet splitting problem and corresponds to a significant tree-level fine tuning. 
On the other hand, orbifold SUSY GUTs suppress proton decay as before, and also solve the doublet triplet splitting by using the boundary conditions to remove the light modes of the colored Higgs triplet whilst retaining the light modes of the SM Higgs. 
The ``cost" of standard supersymmetric grand unification from the perspective of cosmological collider physics is a higher unification scale $M_{\rm GUT}\sim 10^{16}$ GeV, placing it out of reach for the cosmological collider. 
In orbifold SUSY GUTs however, the KK excited GUT partners can appear at the lower compactification scale $M_C \sim 10^{15}$ GeV \cite{Hall1}, as illustrated in \fig{fig:Orbifold coupling unification}. However, naively this is still out of reach given  \eq{eq:HubbleBound} and the Boltzmann suppression in \eq{eq:typicalsignal}. 

Nevertheless, there is a ``chemical potential" extension of cosmological collider physics involving direct couplings between the heavy particles and the inflaton, in which Boltzmann suppression above $H$ can be evaded \cite{CCP14,Wang_2020,Wang:2020ioa,Bodas_2021}. 
In this paper we will develop a variant of the chemical potential mechanism that allows spin-1 GUT excitations to be seen.

As in thermodynamics, the idea of a chemical potential is that it can overcome Boltzmann suppression.
In the inflationary context the mechanism takes advantage of a larger scale than $H$ during inflation, namely the kinetic energy of the inflaton \cite{Planck}
\begin{equation}
    \dot{\phi}_0 \simeq (60 H)^2,
\end{equation}
where $\phi_0(t)$ is the inflationary slow-roll background. 
There exist a class of dimension-5 ``chemical potential" couplings schematically of the form
\begin{equation}
\label{eq:chempotdim5}
    \mathcal{L} \sim \frac{i}{\Lambda}\nabla_\mu\phi J^\mu,
\end{equation}
where $\Lambda \geq 60H$ in order for the inflationary EFT to be in theoretical control, and $J^\mu$ is a current which is quadratic in the heavy field. 
In the inflationary background, \eq{eq:chempotdim5} reduces to a chemical potential coupling $\mathcal{L}\sim i\lambda J^0$.
Such couplings allow for unsuppressed particle production provided $m \leq \lambda$, where 
 \begin{equation}
 \label{eq:lambdadef}
     \lambda \equiv \dot{\phi}_0/\Lambda
\end{equation}     
is the chemical potential. 
Note, the chemical potential can be as large as the kinetic energy scale
\begin{equation}
    \lambda \leq 60H \lesssim 3\times 10^{15} \textrm{ GeV},
\end{equation}
which is above the compactification scale of orbifold SUSY GUTs. 
The chemical potential mechanism both increases the reach, and widens the mass window probed by cosmological collider physics, thereby allowing us to plausibly directly see GUT particles!

% This paper will demonstrate that a chemical potential can be applied to give a large signal for a spin-1 particle coupled to the inflaton.
Our mechanism of chemical potential applies to a massive complex vector field, which for a simple tree-level process must be a SM singlet in order to decay to singlet inflatons.
This precludes its application to $SU(5)$ and $SO(10)$ GUTs for which there is at most a single SM singlet real vector field among the GUT partners. 
However, trinification, another classic unification scenario with the gauge group $G = SU(3)_C\times SU(3)_L \times SU(3)_R/\mathbb{Z}_3$ \cite{GUTDecomposition}, provides a suitable target as it contains several SM singlet GUT partners. 
The realization of the chemical potential in an orbifold trinification scenario is the focus of this paper, while the extension to fully supersymmetric dynamics is left for future work. We expect the central effect of supersymmetry will be to replace the SM gauge-coupling running by the MSSM running, thereby raising the estimate of the compactification scale to $M_C\sim 10^{15}$ GeV. Here we will simply assume $M_C\sim 10^{15}$ GeV as the target for our chemical potential mechanism.

There has been an earlier implementation of chemical potential for a single real spin-$1$ field, in which the enhanced/suppressed ``charges" correspond to the two helicities (that is, the rough analog of our complex field is made from the two real polarizations) \cite{Wang_2020, Wang:2020ioa}. 
In this case the gauge-invariant form of the chemical potential interaction displays a tachyonic instability that is cut off by the spin-$1$ (gauge-noninvariant) mass, requiring some tuning in parameter space to obtain sizeable NG within theoretical control. 
By contrast our chemical potential interaction is not gauge-invariant in form (gauge-invariance is not mandatory because we are dealing with massive spin-$1$ particles) and avoids such tuning. 
Also, the gauge-invariant chemical potential gives rise to NG only at loop-level, whereas our implementation gives rise to NG at tree level with a relatively straightforward analysis.
However, since the gauge-invariant realization requires only one real spin-1 field, it might be more readily applicable to $SU(5)$ and $SO(10)$ GUTs. We leave this to future work. 

% Our mechanism employed a  pair of real spin-$1$ fields to form the complex field whose $U(1)$ rephasing corresponds to the charge enhanced by the chemical potential. 
% This should be compared with an earlier chemical potential mechanism for a single real spin-$1$
% field, in which the enhanced charge corresponds to rephasing the two helicities (that is, the complex field is made from the two real polarizations). 
% In this case the gauge-invariant form of the chemical potential interaction displays a 
% tachyonic instability that is cut off by the spin-$1$ (gauge-noninvariant) mass, requiring some tuning in parameter space to obtain sizeable NG within theoretical control. 
% By contrast our chemical potential interaction is not gauge-invariant in form (gauge-invariance is not mandatory because we are dealing with massive spin-$1$ particles) and avoids such tuning. 
% Also, the gauge-invariant chemical potential gives rise to NG only at loop-level, whereas as we have seen our chemical potential gives rise to NG at tree level with a relatively straightforward analysis.

The paper is organized as follows. In \Sec{sec:prelims} we introduce our conventions and background material. In particular we summarize the spin-$0$ chemical potential which we generalise to spin-$1$. In \Sec{sec:Model} we introduce the model for spin-$1$ chemical potential. A Stueckelberg analysis is then performed to ensure that the massive spin-$1$ EFT is within theoretical control, and the relevant interactions for our NG calculation are isolated.
Section \ref{sec:Calculations} introduces the stationary phase approximation for calculating the NG. The central structure of the calculation is presented, with the full details left for \app{App:StationaryPhase}. 
We then discuss the apparent late-time divergences due to non-derivative inflaton couplings, explaining why they should cancel based on the shift symmetry under which the inflaton and the heavy field transform. A fuller explanation of this cancellation is presented in \app{App:IRDivergences}. We also calculate the corrections to the power spectrum. Requiring this to be subdominant places a bound on the strength of our NG signal.
In \Sec{sec:Results} we discuss observational prospects, a method to determine both the mass of the spin-$1$ particle and the chemical potential, and the possibility of inferring the spin. 
% In section \ref{sec:Results} we discuss the prospects for observing signals of this size, how to determine the mass of the exchanged particle as well as the size of the chemical potential, and finally we discuss the prospect of determining the spin from this (or in fact any other) cosmological collider signal. 
In \Sec{sec:orbifoldcoupling} we study trinification, and show that it contains a candidate complex gauge boson that is uncharged under the SM. 
We show how to implement the chemical potential coupling, but with a fully supersymmetric implementation left for future work. 
In \Sec{sec:Discussion} we end with a summary and a discussion of possible avenues for future research.

%%%%%%%%%%%%%%%%%%%%%%%%%%%%%%%%%%%%%%%%%%%%%%%%

\section{Preliminaries}

\label{sec:prelims}
In this section we establish notational conventions, introduce the in-in formalism for computing cosmological correlators, and give a summary of relevant results from the previous paper on chemical potential \cite{Bodas_2021}, which we are generalizing to the spin-$1$ case. 

\subsection{Conventions}
The mostly plus metric signature ($-,+,+,+$) is used throughout, and the Lagrangian densities are defined with the volume form factored out to facilitate easy integration by parts.
% \begin{equation}
%     S = \int d^4 x\sqrt{-g}\mathcal{L},
% \end{equation}
% where $g$ is the metric determinant. 
During slow-roll inflation, the approximate constancy of the Hubble parameter ($H$) justifies approximating inflationary spacetime as de-Sitter (dS) space:
\begin{equation}
\begin{split}
    ds^2 &= -dt^2 + a^2(t)d\vec{x}^2 \quad {\rm with} \,\,\, a(t) = e^{H t},\\
    & = \frac{1}{(H\eta)^2}\left(-d\eta^2 + d\vec{x}^2\right).
\end{split}
\end{equation}
Here, $\eta = \int dt/a(t)$ is the conformal time with $a(t)$ being the scale factor. In dS spacetime, $\eta = 1/(a H)$. For the sake of convenience, we will set units such that the approximately constant Hubble parameter $H \equiv 1$, restoring $H$ explicitly as needed by dimensional analysis. 

The inflationary spacetime is most simply realized by the dynamics of a scalar ``inflaton" field, $\phi$, when it dominates the energy density of the universe.
This field can be conveniently separated into a slowly-varying homogeneous background, $\phi_0(t)$, which drives inflation and fluctuations, $\delta\phi(\vec{x},t)$, which seed the inhomogeneities of the later universe after inflation.
The fluctuations of the inflaton and the scalar perturbations of the metric, however, are coordinate dependent, with the gauge-invariant combination being the comoving curvature perturbation, $\mathcal{R}$ \cite{Mukhanov:1990me,CPT2}. Since spatial translation is still a good symmetry of the background inflationary dynamics, the correlations of curvature perturbations have the following form,
\begin{equation}
    \langle \mathcal{R}_{\vec{k}_1}\cdots \mathcal{R}_{\vec{k}_n}\rangle = (2\pi)^3 \delta(\vec{k}_1 + \cdots  + \vec{k}_n)\langle \mathcal{R}_{\vec{k}_1}\cdots \mathcal{R}_{\vec{k}_n} \rangle',
\end{equation}
where the prime indicates that the comoving momentum-conserving delta function has been factored out.

The CMB data favors a nearly scale-invariant primordial power spectrum \cite{Planck}
\begin{equation}
    \mathcal{P}_{\mathcal{R}}(k)=\langle\mathcal{R}_{\vec{k}}\mathcal{R}_{-\vec{k}}\rangle' = \frac{1}{(\dot{\phi_0})^2}\frac{1}{2k^3}(k/k_*)^{n_s-1},
\end{equation}
where the scalar spectral index $n_s\approx 0.96$, and the reference comoving momentum scale $k_* \approx 0.05 \rm Mpc^{-1}$. Beyond the power spectrum there will be  non-Gaussian (NG) higher-point correlations in $\mathcal{R}$. 

The focus of this paper is the 3-point correlation function, also called the  primordial bispectrum, $\langle \mathcal{R}_{\vec{k}_1}\mathcal{R}_{\vec{k}_2}\mathcal{R}_{\vec{k}_3}\rangle'$.
% \begin{equation}
%    \cancel{ B(k_1,k_2,k_3) = \langle \mathcal{R}_{\vec{k}_1}\mathcal{R}_{\vec{k}_2}\mathcal{R}_{\vec{k}_3}\rangle'.}
% \end{equation}
 \begin{figure}[t]
\centering
\includegraphics[width=.4\textwidth]{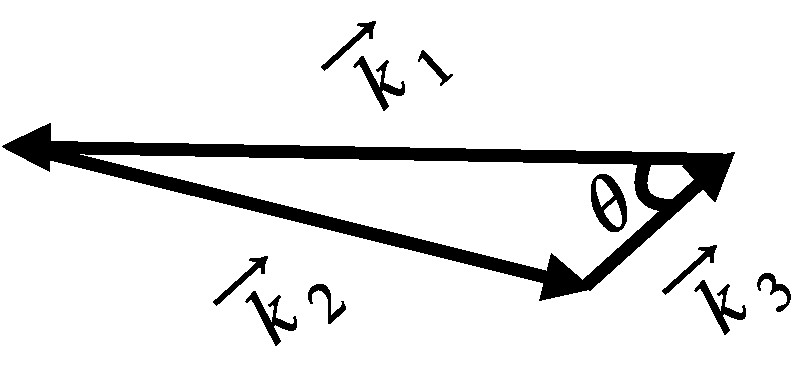}
\qquad
\caption{The shape of the squeezed limit triangle ($k_1 \sim k_2 \gg k_3$) is specified by the degree of squeezing, $p \equiv k_1/k_3$, and the angle between the hard and soft momenta, $\theta$.
\label{fig:squeezedtriangle}}
\end{figure}
It is conventional to characterise NGs relative to the size of the power spectrum \cite{Planck:2019kim}
\begin{equation}
\label{eq:Fdef}
    F(k_1,k_2,k_3) = \frac{5}{6}\frac{\langle \mathcal{R}_{\vec{k}_1}\mathcal{R}_{\vec{k}_2}\mathcal{R}_{\vec{k}_3}\rangle'}{\mathcal{P}_{\mathcal{R}}(k_1)\mathcal{P}_{\mathcal{R}}(k_2)+\mathcal{P}_{\mathcal{R}}(k_1)\mathcal{P}_{\mathcal{R}}(k_3)+\mathcal{P}_{\mathcal{R}}(k_2)\mathcal{P}_{\mathcal{R}}(k_3)}.
\end{equation}
Momentum conservation implies that the three comoving momentum vectors can be arranged to form a triangle. There are two limits of the above expression that are of interest to us. Firstly, the size of NG is typically quoted in terms of a quantity, $f_{\rm NL}$, defined by the equilateral triangle configuration
\begin{equation}
    f_{\rm NL} \equiv F(k,k,k).
\end{equation}
Secondly, the characteristic cosmological collider signal appears in the squeezed limit $k_1\sim k_2\gg k_3$, illustrated in figure \ref{fig:squeezedtriangle}. 

As of now,  primordial NG has not been detected in the CMB, which roughly puts the bounds on $f_{\rm NL}\lesssim \mathcal{O}(5-10)$ \cite{Planck:2019kim} (the exact constraint is shape-dependent.)
However, future datasets are expected to reach higher precision for NGs compared to the CMB. 
For example, large-scale structure surveys (LSS) are projected to be sensitive to $f_{\rm NL}\sim \mathcal{O}(1)$ \cite{LSS1}, while a more futuristic 21-cm tomography could probe NG as low as $f_{\rm NL} \sim \mathcal{O}(10^{-2})$ \cite{21cm1,21cm2}. These datasets would be sensitive to our benchmark signals in section \ref{sec:Calculations}.

% Constraints on NGs are shape dependent, but current bounds are $F_{\rm NL}\sim\mathcal{O}(5-10)$, and LSS is expected to bring the bounds down to $F_{\rm NL}\sim \mathcal{O}(1)$ \cite{LSS1}. Future 21-cm cosmology could probe as low as $F_{\rm NL} \sim \mathcal{O}(10^{-2})$ \cite{21cm1,21cm2}.

Calculation of the bispectrum is performed in the in-in formalism. The expectation value of a Heisenberg picture operator at a time $t$ is given by \cite{CPT1,Weinberg_2005}
\begin{equation}
    \label{eq:ininmasterformula}
    \langle\mathcal{O}_H(t)\rangle = \langle 0 | \bar{T}\left\{e^{i\int^t_{-\infty(1+i\epsilon)}H_I(t')dt'}\right\}\mathcal{O}_I(t)T\left\{e^{-i\int^t_{-\infty(1-i\epsilon)}H_I(t')dt'}\right\}|0\rangle,
\end{equation}
where the subscript $I$ indicates an interaction picture operator and $H_I(t)$ is the  interaction Hamiltonian. The state $|0\rangle$ is the Bunch-Davies ``vacuum" corresponding to the free Minkowski vacuum at asymptotically early times (when each k mode has physical wavenumber $|k\eta|\gg H$), while the $i\epsilon$ prescription projects onto the interacting de-Sitter state as time evolves. The primordial contribution to an expectation value is calculated at the end of inflation (on the reheating surface, for example). Provided the primordial NGs are insensitive to the details of reheating (which is a valid assumption since all the comoving modes that we are interested in exit the horizon before reheating), one can calculate expectation values in exact dS space, with the end of inflation approximated as $t\rightarrow \infty$, or $\eta \rightarrow 0$. In this limit the curvature perturbation is most easily calculated in the spatially flat gauge 
\begin{equation}
\label{eq:ComCurvPert}
    \mathcal{R} = \frac{\delta\phi}{\dot{\phi}_0},
\end{equation}
where the time derivative of the inflaton background $\dot{\phi}_0\approx \rm constant$ during inflation. In the squeezed limit we can simplify \eq{eq:Fdef},
\begin{equation}
\label{eq:Fsqdef}
    F_{\rm sq} \approx \frac{5}{12}\dot{\phi}_0 \frac{\langle\delta\phi_{\vec{k}_1}\delta\phi_{\vec{k}_2}\delta\phi_{\vec{k}_3}\rangle'}{\langle\delta\phi_{\vec{k}_1}\delta\phi_{-\vec{k}_1}\rangle'\langle\delta\phi_{\vec{k}_3}\delta\phi_{-\vec{k}_3}\rangle'} \, ,
\end{equation}where  \eq{eq:ComCurvPert} was used to write the NG in terms of inflaton fluctuations.

\subsection{Recap: chemical potential for spin-0 particles}

The goal of this paper is to extend to spin-1 particles the mechanism of chemical potential that was discussed for spin-0 fields in \Cite{Bodas_2021}. 
We therefore review some of the important lessons from \Cite{Bodas_2021}.
In the minimal cosmological collider, the exponential suppression $\sim e^{-\pi m/H}$ for fields with $m>H$ is a consequence of the fact that the inflating background typically provides energies of order $H$, or intuitively a Hawking temperature $T_{\rm H} = H/(2 \pi)$.
One way to overcome this ``Boltzmann" suppression is to incorporate a ``chemical potential'' that is larger than $T_{\rm H}$, and can allow excitations of particles that are heavier than $H$.
Such a chemical potential can be provided by the kinetic energy of the rolling inflaton.
Coupling the inflaton derivatively to the current of the heavy field can produce a chemical potential-like operator.
For a complex scalar $\chi$ of mass $m$, the Lagrangian looks like
\begin{align}\label{eq:scalar_inflaton-current_coupling}
    % \mathcal{L}_{\chi} \, &\supset \, 
    % \, \left[ (\nabla_{t}+i\frac{\dot{\phi}_0}{\Lambda})\chi \right] \left[ (\nabla_{t}-i\frac{\dot{\phi}_0}{\Lambda})\chi^{\dagger} \right] -|\nabla_{i}\chi|^2- M^2|\chi|^2 +\alpha \chi + h.c  \nonumber\\ 
    \mathcal{L}_{\chi} \, &\supset \, 
    \, \left[ \paren{\nabla_{\mu} + i\frac{\nabla_{\mu} \phi}{\Lambda}}\chi \right] \left[ \paren{\nabla^{\mu} - i\frac{\nabla^{\mu} \phi}{\Lambda}} \chi^{\dagger} \right]- M^2|\chi|^2 +\alpha \chi + h.c  \nonumber\\ 
    & =-|\nabla_{\mu} \chi|^2 - M^2 |\chi|^2 +\underbrace{i\frac{\nabla_{\mu} \phi}{\Lambda} J^{\mu}}_{\rm chem \, pot} + \frac{\nabla_{\mu} \phi \nabla^{\mu} \phi}{\Lambda^{2}} |\chi|^2 + \underbrace{\alpha \chi + h.c}_{{\rm explicit} \,\cancel{U(1)}}  \:.
\end{align}
where $\nabla_{\mu}$ is the diffeomorphism-covariant derivative and $J^{\mu} = i(\chi \nabla^\mu \chi^\dagger - \chi^\dagger \nabla^\mu\chi)$ is the Noether current associated with the global $U(1)$ rephasing symmetry of $\chi$. 
The role of a small explicit $U(1)$ symmetry-breaking term in \eq{eq:scalar_inflaton-current_coupling} will becomes clear shortly.  
After plugging in the background for $\phi$, we see that the Lagrangian contains a charge density operator, $\frac{\dot{\phi}_0}{\Lambda} J^{0} = \lambda J^{0}$, with a chemical potential $\lambda = \frac{\dot{\phi}_0}{\Lambda}$.

The effect of a chemical potential is more apparent in a ``rotated" basis $\chi \mapsto e^{-i\phi/\Lambda}\chi$, where the Lagrangian in \eq{eq:scalar_inflaton-current_coupling} becomes
\begin{align}
    \mathcal{L}_{\chi} \xrightarrow{\rm rotated}& -|\nabla_{\mu} \chi|^2 - M^2 |\chi|^2 + \alpha \chi e^{-i \phi/\Lambda} + h.c. \nonumber \\
    =& -|\nabla_{\mu} \chi|^2 - M^2 |\chi|^2 + \alpha \chi e^{-i \lambda t} e^{-i \delta \phi/\Lambda} + h.c. \, ,
    \label{eq:scalarcoupling}
\end{align}
where we have used $e^{-i\phi/\Lambda} = e^{-i(\phi_0+\delta\phi)/\Lambda} \approx e^{-i\lambda t}e^{-i\delta\phi/\Lambda}$. 
The importance of symmetry-breaking is now manifest; it acts as a ``bath" that allows us to violate $\chi$ number conservation. However, there remains a non-linearly realised symmetry 
\begin{eqnarray}
    &\chi \mapsto e^{ia/\Lambda}\chi,\nonumber \\
    &\delta \phi \mapsto \delta\phi + a,\label{eq:residualsymmetry}
\end{eqnarray}
protecting the flatness of the inflaton slow-roll potential, that ensures correlators are free of late time divergences. 

The coupling in \eq{eq:scalarcoupling} is effectively a linear source term for $\chi$, where the source has frequency $\lambda$. 
This source can now produce particles as heavy as $\lambda$ without any suppression.
This is further illustrated in \Sec{subsec:Calculations_outline}.
The non-Gaussianity due to an exchange of a heavy scalar with $m<\lambda$ was found to have the form
\begin{equation}
    F_{\rm sq} \approx |f_{\rm oscil}(\mu_0,\lambda)|\left(\frac{k_1}{k_3}\right)^{-3/2 + i(\mu_0 -\lambda)} + c.c.
\end{equation}
The signal is unsuppressed for $\mu_0 < \lambda$ as expected, with a potentially observable  $|f_{\rm oscil}| \sim {\cal O}(0.1-1)$ \cite{Bodas_2021}.

The key takeaways from \Cite{Bodas_2021} for implementing the chemical potential are:
\begin{enumerate}
    \item An explicit breaking of the heavy particle number conservation is required,
    \item The chemical potential can be implemented as a phase $e^{\pm i \phi/\Lambda}$ so as to nonlinearly realise the symmetry in \eq{eq:residualsymmetry}.
\end{enumerate}
We will use these simplified principles in our study to generalize the mechanism to complex vector fields.

\section{A Chemical Potential for Tree-Level Spin-1 Exchange}
\label{sec:Model}
\subsection{Chemical potential in the rotated basis}
This section establishes the spin-1 equivalent of \eq{eq:scalarcoupling}. 
Consider a \textit{complex} vector field $A_\mu$ of mass $m$. The $U(1)$ invariant Proca Lagrangian is
\begin{equation}
\label{eq:proca}
\mathcal{L}_{\rm pr} = -\frac{1}{2}|F_{\mu\nu}|^2 - m^2|A_\mu|^2.
\end{equation}
As seen in \eq{eq:scalarcoupling} in the previous section, a chemical potential can be realized as a $\phi$-dependent phase in the $U(1)$ symmetry-breaking term.
The simplest $U(1)$ breaking term coupling $A_\mu$ linearly to the inflaton is naively $\mathcal{L}_{\rm int} = c \nabla_\mu\phi A^\mu e^{-i\phi/\Lambda} + c.c.$
% This section establishes the spin-1 equivalent of equation \eqref{eq:scalarcoupling}. 
% Consider a \textit{complex} vector field $A_\mu$ of mass $m$. The simplest $U(1)$ breaking term coupling $A_\mu$ to inflation is
% \begin{equation}
%     \label{eq:naivelagrangian}
%     \mathcal{L}_{\rm naive} = \mathcal{L}_{\rm pr} + \mathcal{L}_{\rm s.r} + c \nabla_\mu\phi A^\mu e^{-i\phi/\Lambda} + c.c.
% \end{equation}
% where $\mathcal{L}_{\rm pr}$ is the $U(1)$ invariant Proca lagrangian
% \begin{equation}
% \mathcal{L}_{\rm pr} = -\frac{1}{2}|F_{\mu\nu}|^2 - m^2|A_\mu|^2,
% \end{equation}
% and $\mathcal{L}_{\rm s.r}$ is a slow-roll inflationary lagrangian. 
However, this interaction is redundant because integration by parts  yields a term, $\mathcal{L}_{\rm int} = -ic \Lambda (\nabla_\mu A^\mu) e^{-i\phi/\Lambda}$, which is zero in the leading order equations of motion $\nabla_\mu A^\mu=0$. This coupling can therefore be removed by a field redefinition, 
\begin{equation}
    A_\mu \mapsto A_\mu + \frac{c^{*}}{m^2}\nabla_\mu\phi e^{i\phi/\Lambda}.
\end{equation}
The leading non-redundant interaction that is linear in the vector field comes from a dimension-7 operator, \footnote{Such an interaction can arise by integrating out a heavy scalar, $\sigma$, which interacts with the inflaton via two couplings of the form $\mathcal{L}\sim \lbrace \sigma\nabla_\mu\phi A^\mu e^{-i\phi/\Lambda}$, $\sigma (\nabla_\mu\phi)^2 \rbrace$.}
% \begin{equation}
% \label{eq:lagrangian}
%     \mathcal{L} = \mathcal{L}_{\rm pr} + \mathcal{L}_{\rm s.r} + \frac{c}{\Lambda^3}\nabla_\mu\phi A^\mu \nabla_\rho\phi \nabla^\rho\phi e^{-i\phi/\Lambda} + c.c.
% \end{equation}
\begin{equation}
\label{eq:lagrangian}
    \mathcal{L}_{\rm int} = \frac{c}{\Lambda^3}\nabla_\mu\phi A^\mu \nabla_\rho\phi \nabla^\rho\phi e^{-i\phi/\Lambda} + c.c.
\end{equation}

Despite high dimensionality, this interaction is relevant as it contributes to the bispectrum at tree-level, which makes the calculation analytically tractable.
The interaction above, along with the Proca Lagrangian, has the residual nonlinearly realized symmetry,
\begin{eqnarray}
\label{eq:shift1}
    \phi&\mapsto& \phi + \alpha,\nonumber \\
    A_\mu&\mapsto& e^{i\alpha/\Lambda}A_\mu,
    \label{eq:shift2}
\end{eqnarray}  
which protects the flatness of the inflaton potential, and ensures no late-time divergences as in \eq{eq:residualsymmetry}.

%The lagrangian is also related to one in which the inflaton is manifestly derivatively coupled by a field redefinition akin to the one given in the preliminaries. We leave a larger discussion of IR divergences to appendix B, except to mention that every argument we've made has relied on some amount of integration by parts, but in dS space there is a boundary at $\eta\rightarrow 0$. The two lagrangians may therefore differ by boundary localised actions, so to avoid ambiguities we calculate NGs using an approximation method which steers clear of the $\eta\rightarrow 0$ boundary.

\subsection{Expansion parameters}

Self-interactions of the inflaton follow a derivative expansion $\mathcal{L}(\nabla_\mu\phi/\Lambda^2)$, where $\Lambda$ is roughly the scale of the EFT breakdown. Requiring this to be a controlled expansion gives the first constraint
\begin{equation}
\label{eq:Lambda}
    \Lambda \gtrsim (\dot{\phi}_0)^{1/2} \approx 60 H.
\end{equation}
Combining with the definition of the chemical potential in \eq{eq:lambdadef}, \eq{eq:Lambda} implies
\begin{equation}
\label{eq:lambda}
    \lambda \lesssim 60H.
\end{equation}
The strength of massive vector interactions at high energies is in general sensitive to both its mass and couplings.
Recall the propagator (in Minkowski space)
\begin{equation}
    D_{\mu\nu}(p^2)=\frac{-i}{p^2-m^2+i\epsilon}\left(\eta_{\mu\nu}-\frac{p_\mu p_\nu}{m^2}\right), 
\end{equation}
has a term that scales as $1/m^2$, which enhances the interactions of longitudinal modes in the UV. Determining the scale at which massive vector interactions get strong, and thus the EFT breaks down, is most easily done with the Stueckelberg trick. The high-energy behavior of the longitudinal mode can be isolated by restricting to the form \cite{Stueckelberg:1938hvi} (see \cite{Arkani_Hamed_2003} for a review):
\begin{equation}
\label{eq:longmode}
A_\mu \rightarrow \frac{1}{m}\nabla_\mu\pi.
\end{equation}
With this form, the interaction in \eq{eq:lagrangian} looks like
\begin{equation}
    \mathcal{L}_{\rm int} \approx \frac{c}{m\Lambda^3}\nabla_\mu\phi \nabla^\mu\pi \nabla_\rho\phi \nabla^\rho\phi e^{-i\phi/\Lambda}.
\end{equation}
As expected, the interactions of the longitudinal mode are enhanced by a factor of $c E/m$ for momenta of typical size $E$. Requiring that the interactions remain weak up to $E\sim\Lambda$ gives the final constraint
\begin{equation}
\label{eq:EFTc}
    c < \frac{m}{\Lambda}.
\end{equation}

\subsection{Other lower dimension interactions}
While we have stated the leading linear interaction between $\phi$, and $A_\mu$, there are lower dimension interactions that are quadratic in $A_\mu$. We choose a power counting for these couplings consistent with \eq{eq:lagrangian} in which every $A_\mu$ appearing in interactions with $\phi$ is multiplied by $c$. In this way the leading such interactions appear at dimension-6 and are given by
\begin{equation}
    \mathcal{L}_6 \sim \frac{|c|^2}{\Lambda^2} A^\dagger_\mu A^\mu \nabla_\nu\phi\nabla^\nu\phi,\, \frac{|c|^2}{\Lambda^2}A^\dagger_\mu A_\nu \nabla^\mu\phi \nabla^\nu\phi.
\end{equation}
These provide generalised mass terms $\delta m^2 \sim |c|^2\lambda^2$.  However, due to equations \eqref{eq:lambda} and \eqref{eq:EFTc} these are subleading to the mass term in \eqref{eq:proca}, and will be neglected here.

In addition to these higher dimension terms there is the possibility of providing a chemical potential for pairs of vectors within loop level processes
\begin{equation}
    \mathcal{L}_{\rm int} \sim |c^2|\Lambda^2 A_\mu A^\mu e^{-2i\phi/\Lambda} + h.c.
\end{equation}
This is an interesting possibility which we will explore in future work, but in this paper we will focus on the tree-level effect due to the linear coupling. 

\subsection{Interactions}

The interactions can be obtained from \eq{eq:lagrangian} by separating the fields into their homogeneous backgrounds and fluctuations, $\phi(\vec{x},\eta) = \phi_0(\eta)+\delta\phi(\vec{x},\eta)$, $A_\mu(\vec{x},\eta) = \mathcal{A}_\mu(\eta) +  A_\mu(\vec{x},\eta)$. To avoid confusion with the time component of a 4-vector we have called the vector background ``$\mathcal{A}_\mu$", and we have reused ``$A_\mu$" (rather than ``$\delta A_\mu$") for the fluctuation to avoid overly cumbersome notation. 

When the background for the vector is plugged into \eq{eq:lagrangian}, we are only left with inflaton self-interactions. 
While these interactions are important for the evaluation of the full shape of non-Gaussianity, they do not contribute to the non-analytic part as it requires on-shell propagation of a heavy particle.
Therefore, we will not discuss the effects of inflaton self-interactions in this work.
% Such interactions aren't relevant for calculating the non-analytic contribution to NGs because the non-analytic dependence is a signal of the on-shell propagation of a heavy particle. 
The full set of interactions and the expression for the homogeneous background of the vector field can be found in \app{App:Ints}. 
Below we only collect terms that are relevant for the calculation of the non-analytic part of the NG.
For the bispectrum, both mixing and three-point vertices are important.

The quadratic mixing terms between the two fields are as follows,
\begin{equation}
    \mathcal{L}_{\rm mix} = -\mathcal{H}_{\rm mix}= \frac{c}{\Lambda}(-\eta)^{i\lambda}\left\{i\eta\lambda^3A_0 \delta\phi +3\eta^2\lambda^2A_0\delta\phi' - \lambda^2\eta^2 A_i\partial_i\delta\phi\right\}+h.c. \, ,
    \label{eq:Mixing}
\end{equation}
where the prime indicates $' \equiv \partial /\partial \eta$. 

At tree-level, the relevant three-point interactions are those that contain two inflaton fluctuations and one vector fluctuation,
\begin{align}
    \nonumber \mathcal{L}_{\rm A\phi\phi} = -\mathcal{H}_{\rm A\phi\phi} &= \frac{c}{\Lambda^2}(-\eta)^{i\lambda}\left\{\frac{\lambda^3\eta}{2}A_0 \delta\phi^2 - 3i\lambda^2\eta^2A_0\delta\phi\delta\phi'-3\lambda\eta^3A_0(\delta\phi')^2\right.\\   &\hspace{1.3cm}+\lambda\eta^3A_0(\partial_i\delta\phi)^2+i\lambda^2\eta^2A_i\delta\phi\ \partial_i\delta\phi+2\lambda\eta^3A_i\delta\phi'\ \partial_i\delta\phi\bigg\} + h.c.
    \label{eq:threepointinteractions}
\end{align}
In the next section we derive the central features of the NG calculation, while the detailed calculation is performed in \app{App:StationaryPhase}. 

%\mathcal{L}_{A\phi\phi} &=& \frac{c}{\Lambda}\frac{\lambda}{\Lambda}(-\eta)^{i\lambda}\eta\left(2\delta\phi'\nabla_\mu\delta\phi A^\mu + (\nabla_\mu\delta\phi)^2A_0\right) - 2ic(-\eta)^{i\lambda}\frac{\lambda^2}{\Lambda^2}\eta^2\delta\phi\delta\phi'A_0 + h.c.

%%%%%%%%%%%%%%%%%%%%%%%%%%%%%%%%%%%%%%%%%%%%%%%%%%

\section{Non-Gaussianity due to Massive Spin-1 Exchange}
\label{sec:Calculations}

\subsection{The stationary phase approximation}
\label{subsec:Calculations_outline}

\begin{figure}[t]
\centering
\includegraphics[width=0.7\textwidth]{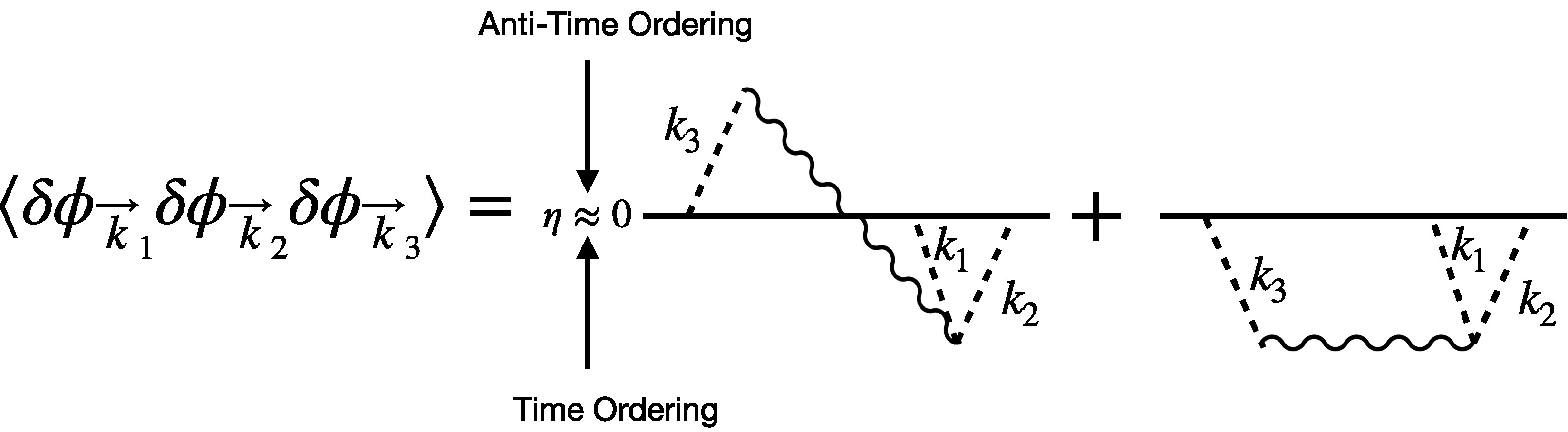}
\qquad
\caption{The tree level contributions to the three-point function due to exchange of $A_\mu$. Vertices above $\eta\sim 0$ come from the expansion of the anti time-ordered exponential in \eq{eq:ininmasterformula}, whilst the vertices below come from the time ordered exponential. The wavy line is the heavy vector, and the dashed lines are inflaton fluctuations.
\label{fig:Bispectrum}}
\end{figure}

The bispectrum can be computed by plugging interactions into the in-in formula of \eq{eq:ininmasterformula},
\begin{equation}   
    \langle \delta\phi^3\rangle|_{\eta \rightarrow 0} = \langle 0 | \left\{\bar{T} e^{i\int^\infty_{-\infty(1+i\epsilon)}H_I(t')dt'}\right\} (\delta\phi^3)|_{\eta \rightarrow 0} \left\{T e^{-i\int^\infty_{-\infty(1-i\epsilon)}H_I(t')dt'}\right\}|0\rangle,
\end{equation}
where we have used a shorthand $\delta \phi^3 \equiv \delta\phi_{\vec{k}_1}\delta\phi_{\vec{k}_2}\delta\phi_{\vec{k}_3}$.
We restrict to the relevant interactions described in \eqs{eq:Mixing}{eq:threepointinteractions} in the previous section, $(\mathcal{H}_{\rm mix}, \, \mathcal{H}_{\rm A\phi\phi}) \subset \mathcal{H}_{I} $.
By expanding the exponential, we can see that the dominant non-analytic contribution occurs at tree-level with two vertices and four types of diagrams,
\begin{align}\label{eq:bisp_I_contris}
    \langle \delta\phi_{\vec{k}_1}\delta\phi_{\vec{k}_2}\delta\phi_{\vec{k}_3}\rangle'_{\eta \rightarrow 0} & = I_{-+}+ I_{+-} + I_{++}+ I_{--} \nonumber \\ 
    &= 2 \textrm{Re}(I_{-+} + I_{++}), 
\end{align}
where the $+/-$ notation indicates whether the mixing and the 3-point vertices are chosen from the time-ordered/anti time-ordered exponential respectively. 
It is easy to check that $I_{+-}$ and $I_{--}$ are the complex conjugates of $I_{-+}$ and $I_{++}$ respectively, leading to the simplification in the second line in \eq{eq:bisp_I_contris}.
The corresponding diagrams are shown in \fig{fig:Bispectrum}.

Let us first focus on the $I_{-+}$ diagram, which is easier to analyse and is useful in understanding the role of the chemical potential. 
Since the time ordering is trivial in this diagram, the calculation factorizes into two time integrals, one at each vertex,
\begin{equation} 
\label{eq:Iminplus}
    I_{-+} = \frac{|c|^2}{\Lambda^3}I_{k_3}^- I_{k_{12}}^+ + c.c. \,.
\end{equation}
Here, $I_{k_3}^-$ corresponds to the integral at the mixing vertex, while $I_{k_{12}}$ corresponds to the integral at the cubic vertex.
These integrals typically contain polynomial functions accompanied by highly oscillatory phase factors, 
\begin{eqnarray}\label{eq:I_factorized}
    I_{k_3}^-&=&\int_{0}^\infty dx f_-(x)e^{ig(x)}, \nonumber\\
    I_{k_{12}}^+&=&\int_0^\infty dx f_+(x) e^{ih(x,p)},
\end{eqnarray}
where a convenient variable of integration, $x \equiv -k_{3}\eta$, is chosen and
\begin{equation}
p \equiv (k_{1}+k_{2})/k_3,
\end{equation}
is the squeezing parameter. 

Let us focus on the oscillatory phases as they turn out to be important in determining the presence (or absence) of exponential suppression.
Firstly, the chemical potential provides a high frequency source at each vertex as seen in \eqs{eq:Mixing}{eq:threepointinteractions},
\begin{align}
    \mathcal{H}_{I} \propto e^{\pm i \lambda t} = e^{\mp i \lambda \log(-\eta)}.
\end{align}
Secondly, the mode functions for the fields also have oscillatory time-dependence (equivalent of $e^{-iEt}$ in Minkowski space),
\begin{align}   
    \delta \phi(\vec{k},\eta) &\propto e^{\mp i k \eta}, \nonumber \\
     A_\mu(\vec{k},\eta) &\propto e^{\mp i\int^\eta \sqrt{(-k\eta')^2+\mu^2}\, \frac{d\eta'}{\eta'}} = e^{\pm i\int^t dt' E_k(t')}\, ,
\label{eq:MF1}
\end{align}
where $\mu = \sqrt{m^2-1/4} \approx m$ for $m\gg H$, and $+(-)$ in the phases corresponds to particle creation (annihilation).
% \begin{equation}   
%     A_\mu(\vec{k},\eta) \propto e^{\mp i\int^\eta \sqrt{(-k\eta')^2+m^2}\frac{d\eta'}{\eta'}} = e^{\pm i\int^t dt' E_k(t')}.
% \label{eq:MF1}
% \end{equation}
The full expressions for the mode functions of $A_{\mu}$ are given by Hankel functions, as described in \app{App:ModeFunctions}. However, for $m \gg H$, they can be approximated by the WKB solutions to a harmonic oscillator with time-dependent frequency $E_{k}(\eta) = \sqrt{(-k\eta)^2 + m^2} = \sqrt{k_{\rm phy}^{2} + m^2} $, which is used in the second line of \eq{eq:MF1}. 
This is essentially the locally Minkowskian energy of a massive particle in the expanding dS background.
Putting these together, the phases in \eq{eq:I_factorized} can be written as 
\begin{eqnarray}
    g(x)&=&-\lambda\ln x + x + \int^x \frac{dx'}{x'}\sqrt{x'^2+\mu^2}, \nonumber \\
    h(x)&=&\lambda\ln x - px - \int^x\frac{dx'}{x'}\sqrt{x'^2+\mu^2}.
    \label{eq:hphase}
\end{eqnarray}

We see that due to the highly oscillatory nature of the exponential in these time-integrals, the contribution is suppressed except where the phase is stationary.
In $I^{-}_{k_3}$ integral, $g'(x) = 0$ occurs when
\begin{align}
    \label{eq:xminuseqn}
    &(-k_3\eta) + \sqrt{(-k_3\eta)^2 + \mu^2} = \pm\lambda \nonumber\\
    &\rightarrow \,\, x_- \equiv (-k_3\eta_-) = \frac{\lambda^2-\mu^2}{2\lambda}.
\end{align}
Similarly, for $I^{+}_{k_{12}}$ integral, $h'(x) = 0$ happens when
% \begin{eqnarray}
% \label{eq:xl}
%     x_- \equiv (-k_3\eta_-) = \frac{\lambda^2-m^2}{2\lambda}.
% \end{eqnarray}
\begin{align}
\label{eq:xpluseqn}
    &(-k_1\eta) + (-k_2\eta) + \sqrt{(-k_3\eta)^2+\mu^2} = \pm\lambda \nonumber \\
    &\rightarrow \,\, x_+ \equiv (-k_3\eta_+) = \frac{\lambda-\mu}{p}.
\end{align}
% \begin{equation}
% \label{eq:xr}
%     x_+ \equiv (-k_3\eta_+) = \frac{\lambda-m}{p},
% \end{equation}
The interpretation is clear: the integrals are dominated by the time when the energy flowing through the vertex is conserved in the local Minkowski frame.
The intermediate heavy particle is therefore produced on-shell, and can propagate over a long distance leading to the characteristic non-analytic signature. 
If the time integrals do not include a region where energy can be conserved modulo $H$ (supplied by the expanding spacetime), putting the heavy particle on-shell will lead to an exponential suppression analogous to Boltzmann suppression with the chemical potential taken into account \cite{Bodas_2021}.
This is also the reason behind the Boltzmann-like suppression $e^{-\pi \mu}$ in the minimal cosmological collider signal, as a stationary point does not exist on the real axis for \eqs{eq:xminuseqn}{eq:xpluseqn} if $\lambda =0$.
The chemical potential, through the factors of $e^{\pm i\lambda t}$, is effectively injecting or removing energy at each vertex, and is therefore crucial for overcoming the standard Boltzmann suppression of cosmological collider signals.

The non-analytic dependence can be pinned down from the phase at the stationary points,
\begin{equation}
    I_{-+}\propto e^{ig(x_-)} e^{ih(x_+)}.
\end{equation}
Since $x_-$ only depends on $\mu$ and $\lambda$, $e^{ig(x_-)}$ only provides an overall phase. However, $x_+$ depends on the degree of squeezing $p$, and $e^{ih(x_+)}$ will contain the non-analytic dependence on $p$. 
The $h(x)$ in \eq{eq:hphase} can be straightforwardly evaluated to be
\begin{equation}\label{eq:phase_at_x+}
    h(x_+) = \lambda \ln x_+ - p x_+ - \sqrt{x_+^2+\mu^2} - \mu\ln\left(\frac{x_+}{\mu^2+\mu\sqrt{x_+^2+\mu^2}}\right).
\end{equation}
In the squeezed limit $p\gg 1$, every term is approximately independent of $p$ besides the two logs, $(\lambda-\mu)\ln(x_+)$. The result therefore has a non-analytic dependence coming from 
\begin{equation}
    e^{ih(x_+)} = p^{i(\mu-\lambda)}e^{i\delta(\mu,\lambda)}.
\end{equation}
Determining the prefactors in the integrals requires considering the polynomial functions in \eq{eq:I_factorized} and performing the Gaussian integral around the stationary point. 
% This is performed in detail in \app{App:StationaryPhase}. 
The normalised bispectrum given in \eq{eq:Fsqdef} has the form,
% The final result \eqref{eq:3pointsq}, and \eqref{eq:foscil} give a NG
\begin{equation}
    F_{\rm sq} \approx  \frac{1}{2}f_{\rm oscil}(\mu,\lambda,\cos\theta)p^{-5/2+i(\mu-\lambda)}+c.c.
\end{equation}

Following the outline described above, the full calculation of the non-analytic component of the bispectrum is performed in \app{App:StationaryPhase}. Here we simply report the final answer for $f_{\rm oscil}$:
\begin{equation}
    f_{\rm oscil}(\mu,\lambda,\cos\theta)= \frac{40\pi|c^2|}{3\sqrt{2}m^2}\dot{\phi}_0\frac{\lambda^3}{\Lambda^3}\sqrt{\frac{\lambda}{\mu}}\frac{C(\mu,\lambda)\left[A(\mu,\lambda)+B(\mu,\lambda)\cos^2\theta\right]}{(\lambda-\mu)(\lambda^2-\mu^2)},
\label{eq:foscilfin}
\end{equation}
with functions $A(\lambda ,\mu)$, $B(\lambda ,\mu)$, and $C(\lambda ,\mu)$ given by
\begin{align}
\label{eq:ABC_expressions}
    A &= \frac{1}{8} \paren{-3 \lambda^4 + 8 \lambda^3 \mu - \lambda^2 (5\mu^2 +2 i \mu +6) -2 \lambda \mu (\mu^2 -2) +2 \mu^2 (\mu^2+i \mu -1)} \nonumber \\
    &\approx  -\frac{3 \lambda^4}{8} \, , \nonumber \\   
    B &= \frac{1}{2} \paren{\frac{1}{2} i (\lambda-\mu)^2 - \frac{1}{4} \lambda (2i -\lambda+\mu)^2} \paren{\sqrt{\frac{(\lambda-\mu)^2}{p^2}+\mu^2} + \frac{i \paren{\frac{2(\lambda-\mu)^2}{p^2}+\mu^2}}{2 \paren{\frac{(\lambda-\mu)^2}{p^2}+\mu^2}}}  \nonumber \\
    &\approx -\frac{\lambda^3 \mu}{8} \, ,  \nonumber \\
    C &=-\frac{3(\lambda^2 - \mu^2)^2}{4 \lambda^2} + i \paren{1-\frac{i(\lambda^2-\mu^2)}{2\lambda}}\paren{\lambda+\frac{1}{2} \sqrt{\frac{(\lambda^2 +\mu^2)^2}{\lambda^2}}+\frac{i(\lambda^4 +\mu^4)}{(\lambda^2 +\mu^2)^2}} \nonumber \\
    &\approx2i\lambda\,,
\end{align}
where we used $\lambda \gg \mu$, and $p>\lambda/\mu$ for the approximations. 
Note that $B$ is smaller than $A$ by a factor of $\mu/\lambda$, which is relevant for the determination of spin as discussed in \Sec{subsec:constraining_spin}. Also in the smaller squeezing regime, $p<\lambda/\mu$, $B$ will have some modest $p$-dependence. Observing the non-analytic signal in this regime will be useful in determining the particles mass, and will be discussed in the next section.

\subsection{Other subdominant diagrams} 
The vector propagator in all four diagrams requires an $A_\mu$ to contract with an $A^\dagger_{\mu}$, which means one vertex gets $e^{i\lambda t}$, whilst the other gets $e^{-i\lambda t}$. 
In the $I_{-+}$ diagram $\lambda$ is either injected or removed at both vertices, which corresponds to the propagation of an anti-particle or a particle, respectively.
We studied the case of anti-particle propagation above, where energy is injected at both vertices, and a stationary point therefore exists in both integrals.
It is then straightforward to conclude that stationary points do not exist for particle propagation, where energy is instead removed from both vertices. 
This diagram (and its complex conjugate in $I_{+-}$) is therefore exponentially suppressed as $\sim e^{-2\pi (\lambda+\mu)}$.
For the $I_{++}$ and $I_{--}$ diagrams, the energy from the chemical potential can only be injected at one vertex, and must be taken away at the other. 
Therefore, there is always one time-integral in which a stationary point does not exist.
This gives a slightly less severe but still exponential suppression of the form $\sim e^{-\pi (\lambda -\mu)}$.
For more detailed discussion, we refer the reader to \Cite{Bodas_2021}.
We will henceforth only focus on the dominant diagram $I_{-+}$.

\subsection{Apparent late-time divergences}
Lastly, we would like to address the apparent divergences in the integrals.
It seems that the inflaton fluctuations in the vertices that come from the expansion of the exponential phase $e^{-i\delta\phi/\Lambda} = 1 - i\frac{\delta\phi}{\Lambda} + \cdots$ are not derivatively coupled.
Non-derivative couplings of the inflaton typically lead to late-time ($\eta \rightarrow 0$) divergences in the integrals,
\begin{equation}
\label{eq:schematic}
    I = \int_{-\infty}^0 \frac{d\eta}{\eta^4}f(\eta)e^{ig(\eta)}.
\end{equation}
% \begin{equation}
%     \mathcal{L}_{\rm int} \sim e^{-i\delta\phi/\Lambda} = 1 - i\frac{\delta\phi}{\Lambda} + \cdots.
% \end{equation}
This is because the inflaton fluctuations become constant at late time (i.e. superhorizon freezing) and $f(\eta) \propto \eta^{3/2}$ coming from the heavy field dilution is not sufficient to nullify the divergence $1/\eta^4$ from the metric factor.
However, the divergences in our calculations are purely artifacts of the rotated basis that we are using. 
The residual symmetry in \eq{eq:shift1} ensures that the late-time diverges cancel when we sum over all Feynman diagrams, while the stationary phase contributions that are coming from early time remain. 
The reader is referred to \app{App:IRDivergences} for more details.

\subsection{Correction to the power spectrum}
\begin{figure}[t]
\centering
\includegraphics[width=0.7\textwidth]{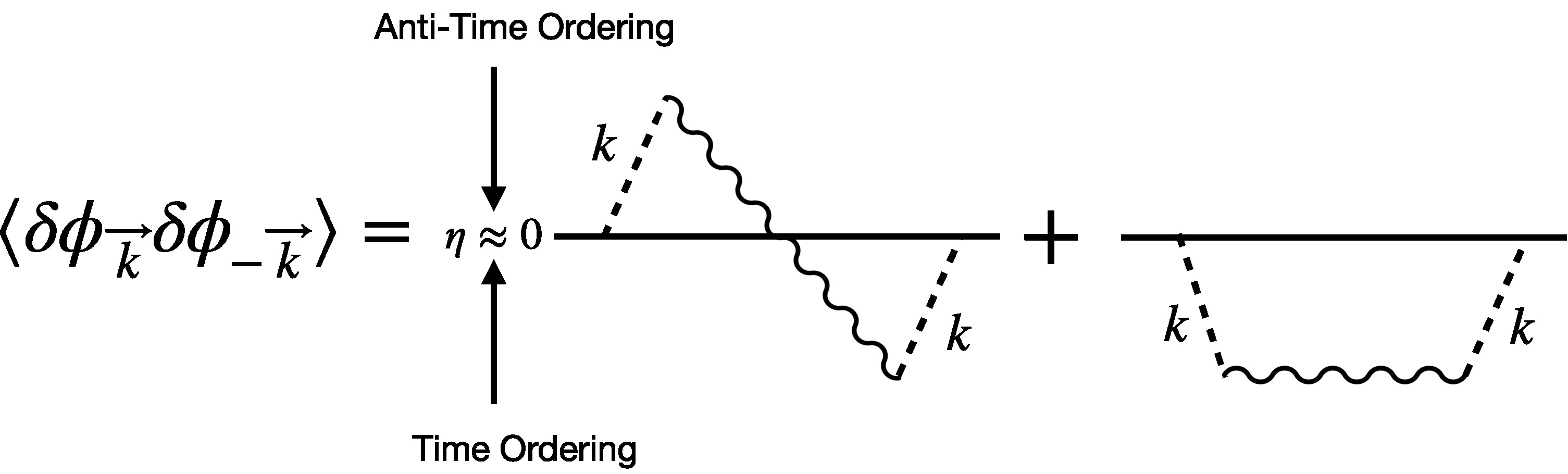}
\qquad
\caption{The tree level contributions to the power spectrum involving the exchange of the vector. The wavy line is the heavy vector, and the dashed lines are inflaton fluctuations. Kinematic considerations show that the $I_{-+}$ diagram on the left is dominant, and the result can be calculcated in the stationary phase approximation, as was done for the NG. 
\label{fig:powerspectrum}}
\end{figure}
Ensuring corrections to the two-point function are small compared to the usual primordial power spectrum provides a further constraint on $|c|$. To estimate the leading correction we can first consider the case $m > \lambda$. While this case is not the focus of this paper, the calculation can be used to infer the constraint in the region of interest $m < \lambda$.   
For $m>\lambda$, the vector can be straightforwardly integrated out by solving the equations of motion for $A_\mu$ in \eq{eq:lagrangian} and ignoring the kinetic term
\begin{equation}
    A^\mu \approx \frac{\bar{c}}{m^2}\frac{1}{\Lambda^3}\nabla^\mu\phi (\nabla_\rho\phi)^2 e^{-i\phi/\Lambda} .
\end{equation}
Plugging back into the Lagrangian renormalizes the purely inflaton pieces. Expanding about the slow-roll background leads to 
\begin{equation}
    \mathcal{L}_{\rm eff} = -\frac{1}{2}\mathcal{Z}(\nabla_\mu\delta\phi)^2 + \mathcal{L}_{\rm int}(\delta\phi),
\end{equation}
where $\mathcal{Z}-1 = \mathcal{O}(1)|c|^2\lambda^4/m^2\Lambda^2$. The fluctuations should be canonically normalised by $\delta\phi \rightarrow \delta\phi/\sqrt{\mathcal{Z}}$. This has an effect on the power spectrum
\begin{equation}
    \label{eq:PowerSpectrumResultmaintext}
    \langle\delta\phi_{\vec{k}} \delta\phi_{-\vec{k}}\rangle' = \frac{1}{\mathcal{Z}}\frac{1}{2k^3} \approx \frac{1}{2k^3}\left[1 + \mathcal{O}(1)\frac{|c|^2\Lambda^2}{m^2}\frac{\lambda^4}{\Lambda^4}\right].
\end{equation}
Since \eq{eq:PowerSpectrumResultmaintext} is scale invariant, sizable corrections to the power spectrum are tolerable. For example, imposing the conservative constraint that the correction is at most 10\% of the free field contribution yields
\begin{equation}
\label{eq:Constraints1}
|c|^2\frac{\Lambda^2}{m^2} < 0.1 \frac{\Lambda^4}{\lambda^4}.
\end{equation}
In the scenario of interest for cosmological collider physics, $m \lesssim \lambda$, there are two regimes: in one regime the heavy particle is significantly off shell, just as in the $m \gtrsim \lambda$ case, and we can expect its contribution to be of similar size. There is also the possibility of the heavy particle being produced on shell, dominated by the $I_{-+}$ diagram for similar kinematic reasons to the NG. The tree level contributions involved the exchange of the heavy vector are shown in figure \ref{fig:powerspectrum}. The $I_{-+}$ diagram factorises and can be readily calculated, yielding
\begin{equation}
    I_{-+} = \mathcal{O}(1) \frac{1}{2k^3}\frac{|c|^2}{m^2\Lambda^2}\frac{ \lambda^5}{(\lambda^2-\mu^2)^2}|C(\mu,\lambda)|^2 \,\, \xrightarrow{\tiny \mu \ll \lambda} \, \mathcal{O}(1) \frac{1}{2k^3}\frac{|c|^2}{m^2\Lambda^2} 4\lambda^3,
\end{equation}
where $C(\mu, \lambda)$ is given in \eq{eq:ABC_expressions}.
Comparing with \eq{eq:PowerSpectrumResultmaintext}, the on-shell contribution is subdominant for masses $\mu < \lambda/2$. Therefore, we obtain the same conservative constraint that we derived from $m \gtrsim \lambda$, namely equation \eq{eq:Constraints1}.

\section{Signal Strength and Key Features}
\label{sec:Results}
\subsection{Signal size}

\begin{figure}[t]
\centering
\includegraphics[width=.6\textwidth]{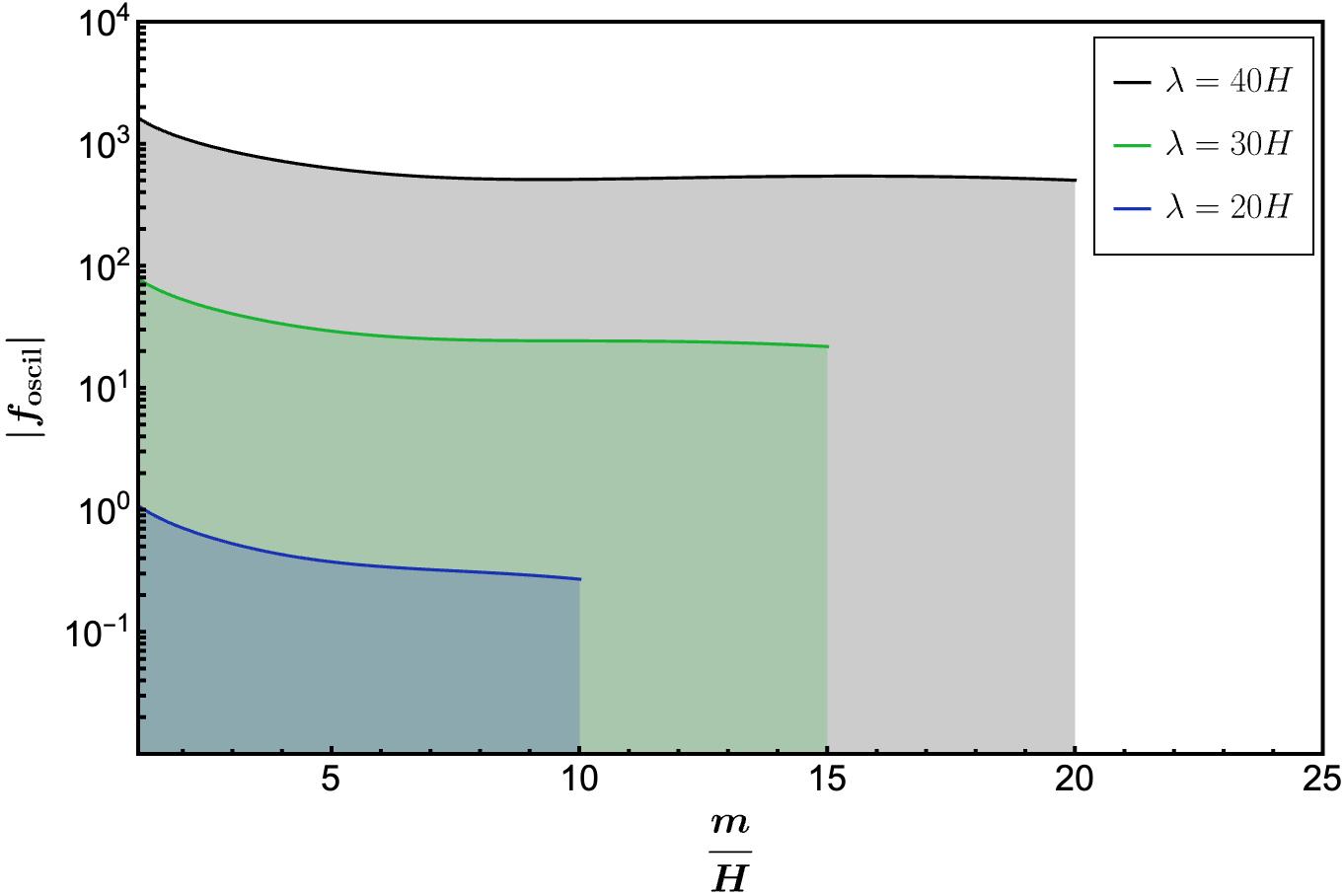}
\qquad
\caption{A plot of the maximum possible signal size $|f_{\rm oscil}|$ within theoretical control as a function of the mass $m$ of the heavy vector field in inflationary Hubble units. We have fixed $\cos\theta=0.5$. For $\mu \lesssim \lambda/2$, the signal strength is within the sensitivity of future LSS and 21-cm experiments.}
\label{fig:foscil}
\end{figure}

The size of the signal is characterized by the amplitude $|f_{\rm oscil}|$ in \eq{eq:foscilfin}. In \fig{fig:foscil} we plot the size of the signal $|f_{\rm oscil}|$ for the largest value of $|c|/m$ consistent with \eqs{eq:EFTc}{eq:Constraints1}.
We also restrict to values of $\lambda \leq 40 H$, in accord with the constraint \eq{eq:lambda} from the requirement of  perturbative control. 

As expected, the exponential suppression is nullified and $|f_{\rm oscil}|$ is enhanced for $\mu \lesssim \lambda$.
% for masses up to the chemical potential. 
For heavier masses beyond the chemical potential, the nonanalyticity in the NG drops exponentially as the on-shell production becomes inefficient. 
The increase in the maximum allowed value of the coupling $c$ with mass is compensated by the $1/m^2$ factor in the NG, as seen in \eq{eq:foscilfin}, which leads to roughly flat profiles in \fig{fig:foscil}. 
% growth of coupling with mass is compensated for by the $1/m^2$ in the result, leading to a shape that is roughly flat. 
% the cutoff is restricted by $\dot{\phi}_0/\Lambda^2 \leq 1/2$, leading to 
We see that depending on the value of the chemical potential, a signal size of 
\begin{equation}
    |f_{\rm oscil}| \sim \mathcal{O}(1-10^3)
\end{equation}
is theoretically possible for a wide range of masses. 
A large portion of this signal space is not excluded by current data, whilst still being able to be probed by future large-scale structure and 21-cm experiments \cite{LSS1,21cm1,21cm2}. 
Note that the stationary phase approximation should not be trusted for values of $\mu$ close to $\lambda$, since in this limit the stationary points are pushed to late times, $x_-,x_+\rightarrow 0$, and the non-analyticity becomes approximately analytic. Therefore, the non-analytic signal would not be separable from the background analytic part leftover after the cancellation of IR divergences and that coming from the inflaton self-interactions.
% , and there is no clean separation between the non-analytic signal and the analytic terms leftover after the cancellation of IR divergences. 
For this reason, we restrict to regions $\mu < \lambda/2$ in figure \ref{fig:foscil}, with the largest mass $m \sim 20H \sim 10^{15} \, {\rm GeV}$.
Even with this conservative choice, the chemical potential allows us to reach the typical mass of the KK gauge bosons in a SUSY orbifold GUT scenario, to be discussed in section \ref{sec:orbifoldcoupling}.

\subsection{Inferring the mass and the chemical potential}
\label{subsec:massandchempot}
\begin{figure}[t]
\centering
\includegraphics[width=.6\textwidth]{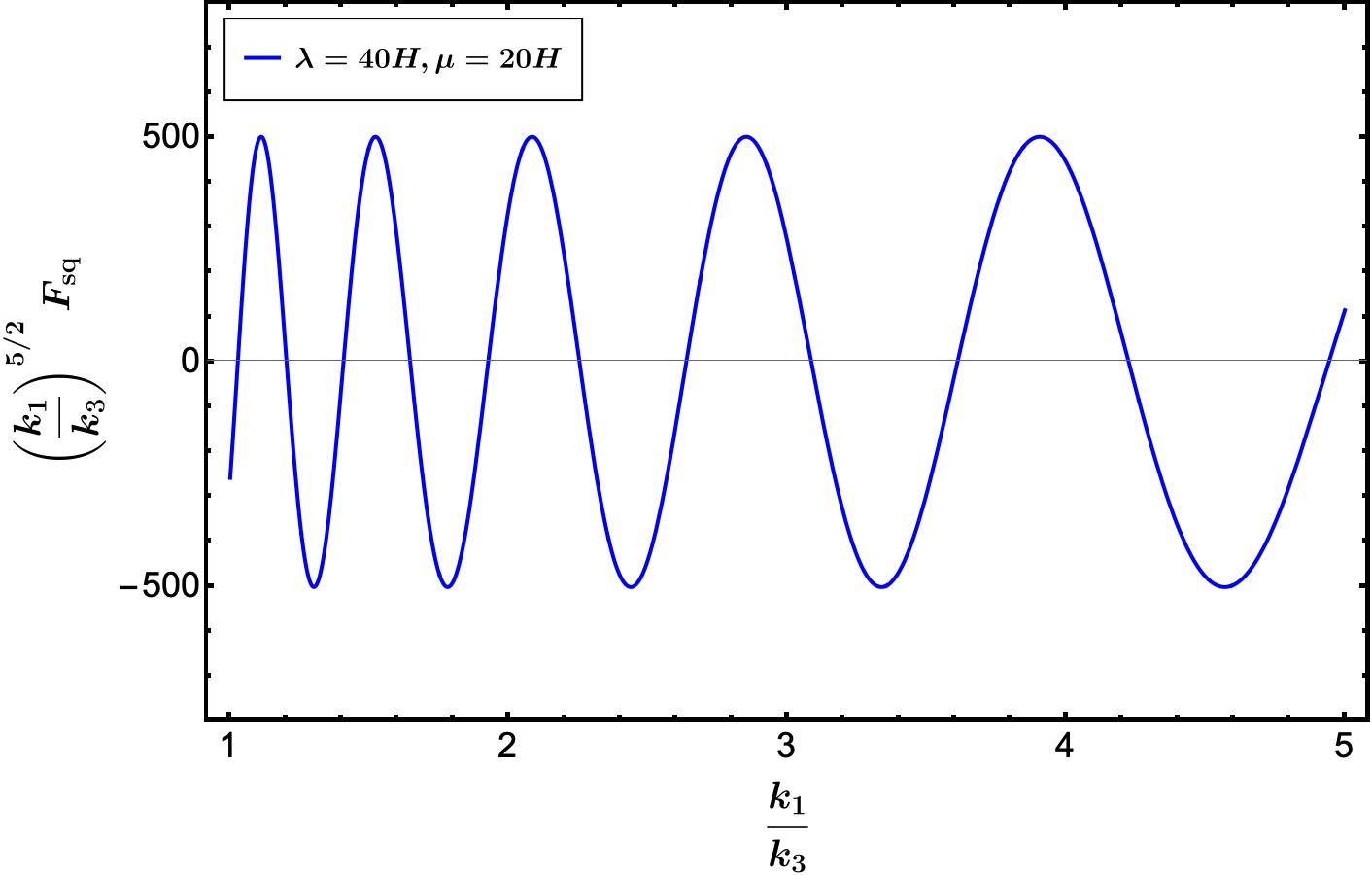}
\qquad
\caption{A plot of the non analytic dependence of $F_{\rm sq}$ as a function of the squeezing $(k_1/k_3)$ with $\cos\theta=1/2$. A measurement of the frequency of oscillation will allow a determination of the combination $\mu-\lambda$.
\label{fig:Fsqueezed}}
\end{figure}

As seen in \eq{eq:foscilfin}, the non-analytic dependence in the squeezed limit has the form
\begin{equation}
    F_{\rm sq} \propto p^{i(\mu-\lambda)},
\end{equation}
which allows for an observation of the combination $(\mu-\lambda)$. Figure \ref{fig:Fsqueezed} shows the non analytic dependence on squeezing for $\mu = 20H$, with $\lambda = 40H$. 

However, this form is correct only in the very squeezed limit.
To see this, let us see more carefully how the stationary point at $k_{12}$-vertex ($x_+$) changes with squeezing.
Using $m \approx \mu$, \eq{eq:xpluseqn}  can be solved to get
\begin{equation}
    x_+ = \frac{p\lambda}{p^2-1}-\frac{\sqrt{\lambda^2+p^2\mu^2-\mu^2}}{p^2-1}.
\end{equation}
At large squeezing such that $p > \lambda/\mu$, all but the second term in the square root can be dropped.
% The squeezed limit corresponds to $p > \lambda/\mu$, in which case all but the second term in the square root can be dropped. 
On the other hand, in the regime of modest squeezing with $p < \lambda/\mu$, all but the first term can be neglected. 
Therefore,
\begin{align}
    x_+ \approx
    \begin{cases}
        \frac{(\lambda - \mu)}{p} \, , &  \, p>\lambda/\mu  \\
        \frac{\lambda}{1+p} \, , &  \, p<\lambda/\mu .
    \end{cases}
\end{align}
% \begin{equation}
%     x_+ \approx \frac{\lambda}{1+p}.
% \end{equation}
Since $p>2$, such a regime can only exist if $\lambda > 2\mu$, which we are restricting to anyway. 

This change in the expression of the stationary point at $k_{12}$-vertex in different regions of the squeezing parameter $p$ has a direct effect on the imaginary exponent in $F_{\rm sq}$.
To see this, let us zoom in on the non-analytic dependence that comes from the phase in \eq{eq:phase_at_x+},
\begin{equation}
    h(x_+) = \lambda \ln x_+ - p x_+ - \sqrt{x_+^2+\mu^2} - \mu\ln\left(\frac{x_+}{\mu^2+\mu\sqrt{x_+^2+\mu^2}}\right).
\end{equation}
For $p \gg \lambda/\mu$ and $\lambda>\mu$, $x_+ \ll \mu$, and the phase simplifies to $(\lambda-\mu)\ln(x_+)$. However, in the modest squeezing regime, $x_+ \gg \mu$, and  the second log term in the above expression becomes approximately independent of $p$. The non-analytic dependence in this case comes entirely from the first log term, which depends only on $\lambda$. Therefore, the non-analyticity in these two regimes goes as 
\begin{equation}
\label{eq:smallsq}
    F_{\rm sq} \propto
    \begin{cases}
         p^{i (\mu-\lambda)} \, , &  \, p>\lambda/\mu  \\
         (1+p)^{i \lambda} \, , &  \, p<\lambda/\mu .
    \end{cases}
\end{equation}
% \begin{equation}
% \label{eq:smallsq}
%     e^{ih(x_+)} \sim e^{i\lambda\ln(x_+)} \sim \left(1+p\right)^{-i\lambda},
% \end{equation}
Observation of this change in the non-analytic exponent as we change the amount of squeezing can be used to infer both the mass and the chemical potential, paralleling the results of spin-0 case from \Cite{Bodas_2021}.

\subsection{Constraining the spin}\label{subsec:constraining_spin}

In \Cite{arkanihamed2015cosmological} it was claimed that NG due to spin-$j$ exchange would have a characteristic dependence on the shape of the momentum-conservation triangle. They used the minimal coupling
\begin{equation}
    \mathcal{L} \sim (\nabla^{\mu_1}\phi) \cdots (\nabla^{\mu_j}\phi) A_{\mu_1 \cdots \mu_j},
\end{equation}
leading to a signal of the form
\begin{equation}
    F_{\rm sq} \propto P_j(\cos\theta)\left(\frac{k_1}{k_3}\right)^{-3/2+i\mu} + c.c. \, ,
\end{equation}
where $P_{j}$ is the Legendre polynomial of a degree that is determined by the spin $j$. 
Observing the signal as a function of $\cos\theta$ would then allow determination of the spin. 

However, it was pointed out in \Cite{Lee_2016} that for odd spin particles, a cancellation between diagrams leads to a more severe dilution factor of $(k_1/k_3)^{-5/2}$. In \cite{Kumar:2017ecc,Kumar:2018jxz} the authors found a $\sin^2\theta$ dependence for NG mediated by spin-1, by considering a minimal higher dimension interaction. 
Here, because of the chemical potential, we find a different shape dependence for spin-1 of the form
\begin{equation}
    F_{\rm sq} \sim \left[\mathcal{F}_1(\mu,\lambda) + \mathcal{F}_2(\mu,\lambda)\cos^2\theta\right]\left(\frac{k_1}{k_3}\right)^{-5/2+i(\mu-\lambda)},
\end{equation}
where $\mathcal{F}_i(\mu,\lambda)$ can be read off from \eq{eq:foscilfin}.
% \abedit{We find that the part with the angular dependence $\mathcal{A}_2$ is $1/\lambda$ smaller than the one without.}

It appears that an observation of the dilution will indicate the exchange of an odd spin particle, but it may be hard to pin down the exact spin since there is no general form for the shape dependence. 
Further investigation is therefore needed to know if the shape dependence can unambiguously determine the spin in cosmological collider physics. It turns out that in our model, $\mathcal{F}_{2}$ is smaller than $\mathcal{F}_{1}$ by a factor of $\mu/\lambda$, so it may be difficult to observe the $\theta$ dependence. However, an observation of any dependence on $\cos\theta$ for the non-analytic signal will be a smoking gun of non-zero spin.
%Check this last sentence is true

\section{Embedding the Chemical Potential within Orbifold Unification}
\label{sec:orbifoldcoupling}
\subsection{Trinification}

Implementing a coupling of the form in equation \eq{eq:lagrangian} requires the vector be a standard model (SM) gauge singlet because the inflaton is uncharged under the SM. 
The simplest $SU(5)$, $SO(10)$ GUTs do not include a complex vector SM singlet. 
However, the classic trinification\footnote{Note that all three unified groups mentioned are subgroups of $E_6$} scenario with a gauge group \cite{GUTDecomposition}
\begin{equation}
    G_{\rm GUT} = SU(3)_C\times SU(3)_L\times SU(3)_R/\mathbb{Z}_3,
\end{equation}
breaking to $G_{\rm SM} = SU(3)_C\times SU(2)_L \times U(1)_Y$ contains several neutral heavy gauge bosons. The discrete $\mathbb{Z}_3$ exchanges the three $SU(3)$ groups ensuring a single gauge coupling.

The symmetry breaking pattern for the gauge bosons is as follows
\begin{eqnarray}
    (8,1,1) &&\rightarrow (8,1)_0,\nonumber \\
    (1,8,1) &&\rightarrow (1,3)_0\oplus(1,2)_{1/2}\oplus (1,2)_{-1/2}\oplus (1,1)_0,\nonumber\\
    (1,1,8) &&\rightarrow 4(1,1)_0 \oplus 2(1,1)_1 \oplus 2(1,1)_{-1},
\end{eqnarray}
where the numbers indicate the dimension of the representations under the GUT (left) or SM gauge group (right) with the subscripts indicating the hypercharge. Of the 5 SM singlets one linear combination is the hypercharge gauge boson, leaving 4 real, or 2 complex massive vectors as candidates for a cosmological collider signal.

\subsection{Realising the chemical potential coupling}

The coupling in \eq{eq:lagrangian} can be realised in the context of orbifold GUTs within 5-dimensional effective field theory. 
\begin{figure}[t]
\centering
\includegraphics[width=.5\textwidth]{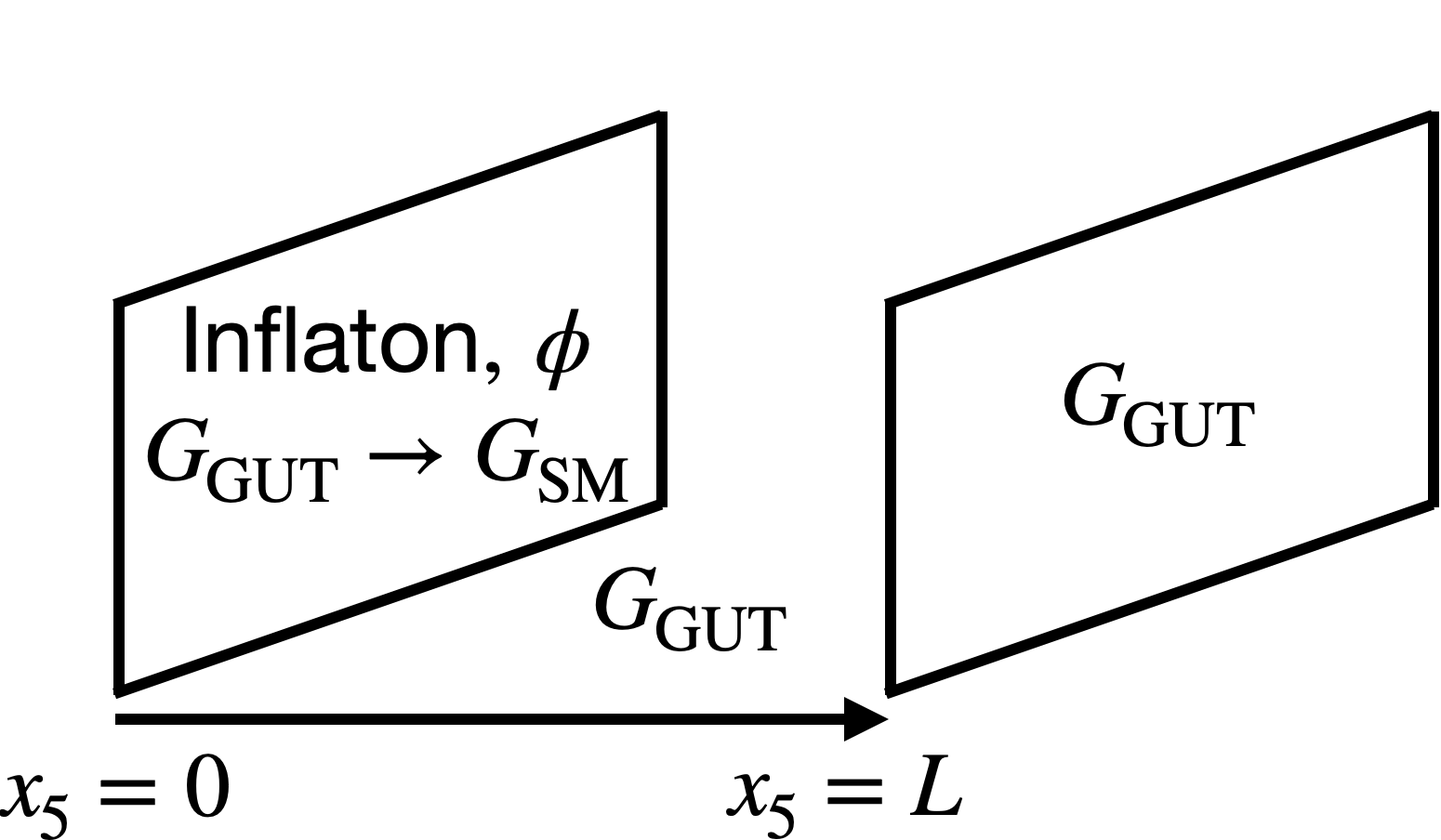}
\qquad
\caption{In orbifold GUTs the symmetry breaking occurs via boundary condition in an extra dimension. We consider the gauge fields to live on an interval of length $L$. We must impose either Dirichlet or Neumann BCs at either end of the interval. Imposing Dirichlet BCs will remove the zero mode, explicitly breaking the gauge invariance in the IR for that particular generator. Neumann BCs at both ends will leave a zero mode, such that after integrating out the extra dimension, the IR theory respects the gauge invariance for that generator.
\label{fig:orbifold}}
\end{figure}
In orbifold GUTs, the symmetry breaking, illustrated in \fig{fig:orbifold}, is achieved by choice of boundary conditions (BC) on one or more (3+1) dimensional boundaries of a higher dimensional ``bulk". The full GUT gauge symmetry is respected by the bulk action, but only the SM gauge invariance is respected on the SM boundary at $x_5=0$. The inflaton is a purely 4D field localised on the boundary at $x_5=0$ \cite{Kumar:2018jxz}. 

The orbifold GUT symmetry breaking from trinification down to the standard model gauge group was studied in \cite{Carone:2004rp, Carone:2005rk}. On the SM boundary, we impose the SM-gauge-invariant conditions
\begin{eqnarray}
\nabla_5 A_\mu^{\rm SM}\vert_{x_5=0}&=&0,\label{eq:SMBC}\\
A_\mu^{\rm \,\cancel{\rm SM}}\vert_{x_5=0}&=&0,
\label{eq:GUTBC}
\end{eqnarray}
where $\cancel{\rm SM}$ indicates broken GUT generators. 
The boundary at $x_5=L$ respects the full GUT gauge invariance, and Neumann BCs are imposed on all generators
\begin{equation}
\label{eq:GUTbrane}
    \nabla_5 A_\mu \rvert_{x_5=L}=0.
\end{equation}
After performing a KK decomposition, the BCs in \eqs{eq:SMBC}{eq:GUTbrane} ensure the existence of a zero mode for the SM gauge bosons, whereas \eq{eq:GUTBC} removes the zero mode for the gauge bosons of the broken generators, leaving only the massive KK modes. 

Let us denote by $Y_\mu = (A_{1}^{\,\cancel{\rm SM}}+i A_{2}^{\,\cancel{\rm SM}})_{\mu}$ one of the candidate complex vectors corresponding to two broken generators that are also SM singlets. The chemical potential coupling can then be implemented via the interaction:
\begin{eqnarray}
    S_{\rm boundary}&&\supset \frac{c_1}{\Lambda_{5D}^{4.5}}\int d^4 x  \nabla_\mu\phi(\nabla_\rho\phi\nabla^\rho\phi) e^{-i\phi/\Lambda} \nabla_5 Y_\mu\vert_{x_5=0},
\end{eqnarray}
where the 5D effective gauge theory strong coupling scale $\Lambda_{5D} \sim 16\pi^2/ N g_5^2$ with $g_5$ the 5D gauge coupling, and $N=3$ are the number of colors of the various trinification groups. Notice that since $Y_\mu$ is a SM gauge singlet the coupling respects the SM gauge symmetry, but not the gauge symmetry associated with the broken generators, which is allowed for terms localised to the boundary at $x_5=0$.

The 5D gauge field KK-decomposes as
\begin{equation}
    Y_\mu(x,x_5) = \sqrt{\frac{2}{L}}\sum\limits_{n=1}^\infty\sin\left(\frac{(n-\frac{1}{2})\pi x_5}{L}\right) Y_\mu^{(n)}(x).
\end{equation}
We will focus on the first KK mode with $n=1$.
The coupling in \eq{eq:lagrangian} can therefore be identified with
\begin{equation}
    c = \frac{c_1 \pi}{\sqrt{2}} \left(\frac{\Lambda}{\Lambda_{5D}}\right)^{3/2}(\Lambda L)^{-3/2},
    \label{eq:couplingsize}
\end{equation}
where $\Lambda$ is the cutoff of the inflationary EFT which must be before the 5D gauge theory cutoff $\Lambda \leq \Lambda_{5D}$.

In the extra-dimensional framework the $5D$ gauge coupling, $g_5$, is non-renormalisable. Below the compactification scale $M_{C} = \pi/L$ we match on to the SM or MSSM effective theory. Tree level matching predicts gauge coupling unification at, $M_C$
\begin{equation}
\label{eq:TreeGauge}
    g_i(M_{C}) = \frac{g_5}{\sqrt{L}},
\end{equation}
where $g_{i=1,2,3}$ are the SM gauge couplings. 

% We don't observe such a unification of gauge couplings in experiments, but if we run the SM gauge couplings up to high energies they come crudely close to unifying around the scale $M_{\rm GUT} \sim 10^{14}$ GeV. The 1-loop matching will give corrections to \eq{eq:TreeGauge}, which depends on the index $i$, and we can match the imperfect unification observed in the real world with a KK scale $M_{\rm KK} \sim 5 \times 10^{13}$ GeV and large corrections to \eq{eq:TreeGauge} \cite{Kumar:2018jxz}.
% In the MSSM, however, the gauge couplings come much closer to unifying than in the non-supersymmetric case. In the SUSY scenario, only small corrections to equation \eq{eq:TreeGauge} are required to fit the data. A detailed accounting of the corrections to \eq{eq:TreeGauge} was given in \cite{Hall1,Hall2,Hall3,Hall4}, which points to a KK scale $M_{\rm KK}\sim 10^{15}$ GeV in the SUSY case.
% This promising case of SUSY GUT can be brought within the reach of cosmological collider by our chemical potential mechanism. 

If we run SM gauge couplings up to high energies (where hypercharge is given its usual GUT normalization), they unify crudely around the scale $M_{\rm GUT} \sim 10^{14}$ GeV. 
The discrepancy may then be plausibly accounted for by loop level and SM brane-localised corrections to \eq{eq:TreeGauge} that depend on the index $i$. A unification can be achieved with a compactification scale of $M_{C} \sim 5 \times 10^{13}$ GeV. As the reader may have noticed, this scale is very close to the inflationary Hubble parameter in high-scale inflation from \eq{eq:HubbleBound}. The cosmological collider prospects for such orbifold GUTs were explored in \Cite{Kumar:2018jxz}. One downside of this setup however is that the corrections must be quite large since the unification in the SM is imperfect. 

This issue is dramatically improved in the minimal supersymmetric standard model (MSSM) as the running gauge couplings come much closer to unifying than in the non-supersymmetric case, and SUSY helps to solve the gauge hierarchy problem between the unification and electroweak scales. In the SUSY scenario, only small corrections to equation \eq{eq:TreeGauge} are sufficient to fit the data to the unification hypothesis. A detailed accounting of corrections to \eq{eq:TreeGauge} was given in \cite{Hall1,Carone:2004rp,Carone:2005rk}, which points to a compactification scale $M_{C}\sim 2\times 10^{15}$ GeV in the SUSY case. 
Interestingly, as pointed out in \Cite{Carone:2004rp}, this particular value of the compactification scale leads to the gauge couplings unifying at the 5D Planck scale. The UV completion of such a model may unify all fundamental forces including gravity. While the compactification scale here is higher than the Hubble scale of inflation $H$, it can still be brought within the reach of cosmological collider discovery via our chemical potential mechanism. 

The full model-building of an orbifold GUT coupled to the inflaton with our chemical potential  within a supersymmetric (supergravity) framework is left to future work. But we expect this to be readily achieved, with the central relevant features already described above.
For the sake of simplification, we have also assumed that only the first KK mode is below the chemical potential scale, i.e. $m_n < \lambda$ only holds for $n=1$. Here $m_n$ is the mass of the $n$th KK mode ($Y_\mu^{(n)}$),
\begin{equation}
    m_n = \left(n-\frac{1}{2}\right) M_{C}.
\end{equation}
Notice that the analysis of \Cite{Carone:2004rp} points to a mass for the first KK mode of $m_1 \sim  10^{15}$ GeV. Combined with the restriction $m_1 \lesssim \lambda/2$ to trust the stationary phase calculation this amounts to requiring $m_1 \in(\lambda/3,\lambda/2)$. The analysis of the previous sections now hold for $Y_\mu^{(1)}$. It would be interesting for future work to consider the case where more than one of the lowest lying KK modes are accessible, $m_n < \lambda$ for more than one $n$. 

\section{Discussion}
\label{sec:Discussion}
In this paper, we have demonstrated that a chemical potential mechanism can be given to a complex spin-1 particle coupled to the inflaton while maintaining EFT control, allowing its propagation to be detected at tree-level. 
In line with the results of \Cite{Bodas_2021} for spin-$0$, we find that the chemical potential can lead to non-Boltzmann-suppressed signals in non-Gaussianities (NGs) for masses up to the chemical potential, which can be as large as $60H$. 
The size of the signal can be within the reach of future experiments in a large part of the parameter space.
% We also found that future experiments will be sensitive to signals sizes th possible in this context. 

% It was also demonstrated that a chemical potential can be given to a vector at tree level, avoiding the loop suppression of previous models. 

With the extended reach for single-production of massive spin-$1$ particles offered by our chemical potential mechanism, the heavy gauge bosons of orbifold-trinification can be targeted.
% An application of this mechanism to detecting orbifold GUTs was then discussed. 
While this also includes the more motivated supersymmetric unification, which has a higher GUT scale compared to the non-supersymmetric case, further model-building is required to realize inflation and chemical potential within full supersymmetric dynamics.
% In the non-supersymmetric case, the results of earlier sections are applicable, but further model building is required to apply the mechanism in a supersymmetric context.
While in this paper we have focused on a chemical potential mechanism that allows single-production (therefore at tree level) of heavy complex fields, we saw in section \ref{sec:Model} that there is also a $U(1)$-breaking term quadratic in the complex vector field which would give rise to pair-production (at loop level). 
Like the inflaton, the pair would have to be a SM-singlet, but now each member of the pair may be SM-charged. We leave this exploration to future work.

We see that an observation of the non-analyticity at intermediate and large squeezing can be used to determine both the mass and chemical potential.
% the quantity $\mu-\lambda$. We also found that seeing the non-analyticity at intermediate squeezing, can be used to infer the two quantities individually. 
Our results are also consistent with the general expectation that NGs of odd-spin heavy particle exchange come with a dilution factor of $(k_1/k_3)^{-5/2}$ \cite{Lee_2016}. It would therefore be possible to say that a cosmological collider signal is due to exchange of a particle with odd spin, but we found a dependence on the shape of the triangle that seems to suggest it may not be possible to say for certain the particle is spin-1. In future work it would be interesting to explore the possibility of determining the spin of a particle observed in cosmological collider physics. In addition to this, the chemical potential mechanism seems generalisable to higher spins, and we hope to apply this methodology to search for higher spin states in the future, including those that might arise in superstring theory.

%Although the realisation of the chemical potential has been discussed in the context of orbifolds, a full model building requires such a coupling to be inside a superpotential. In addition to this, a SUSY orbifold GUT must be successfully coupled to a supersymmetric realisation of inflation. The full model building has therefore been left for future work. In addition to this, our mechanism can be generalised to any model with a spin-1 particle more massive than Hubble. In particular, overcoming the Boltzmann suppression can greatly enhance the size of signals in non supersymmetric GUTs. In the future we will explore the generalisation of our work to orbifold GUTs as well. 

%%%%%%%%%%%%%%%%%%%%%%%%%%%%%%%%%%%%%%%%%%%%%%%%%%%%%%%
\section*{Acknowledgments}
We would like to thank Soubhik Kumar for helpful discussions.
The authors were supported by NSF grant PHY-2210361 and the Maryland Center for Fundamental Physics. In addition, AB was supported by DOE grant DE-SC-0013642 and by Fermi Research Alliance, LLC under Contract No. DE-AC02-07CH11359 with the U.S. Department of Energy, Office of Science, Office of High Energy Physics.

%%%%%%%%%%%%%%%%%%%%%%%%%%%%%%%%%%%%%%%%%%%%%%%%%%%%%%%

\appendix

\section{Mode functions}
\label{App:ModeFunctions}
The mode functions for the inflaton and a massive spin-1 particle in dS space are given below (see \Cite{Lee_2016} for a full derivation). After quantisation, the Fourier transformed inflaton fluctuation is
\begin{equation}
    \delta\phi_{\vec{k}}(\eta) = \bar{u}_k(\eta)c_{-\vec{k}}+u_k(\eta)c^\dagger_{\vec{k}} \,,
\end{equation}
where $c^\dagger, \,c$ are the creation and annihilation operators, respectively. The mode function is given by
\begin{equation}
\label{eq:scalarmodefunctions}
    u_k(\eta) = \frac{1}{\sqrt{2k^3}}(1-ik\eta)e^{ik\eta},
\end{equation}
where $k = |\vec{k}|$. 

For the vector, we focus on the longitudinal mode, which is the only polarization that is affected by the chemical potential in our implementation.
The longitudinal mode of a complex vector is quantized as
\begin{equation}
    A^L_\mu(\vec{k},\eta) = \bar{\sigma}_\mu(k,\eta)\hat{a}_{-\vec{k}} + \sigma_\mu(k,\eta)\hat{b}^\dagger_{\vec{k}},
\end{equation}
where $\hat{a}_{\vec{k}},\hat{b}_{\vec{k}}$ are the annihilation operators for the longitudinal mode of the particle, antiparticle respectively. The bar indicates complex conjugation. For the longitudinal mode
\begin{equation}
        \bar{\sigma}_\mu= \left(\bar{\sigma}_0(k,\eta),\bar{\sigma}_1(k,\eta)\vec{\epsilon}_{\vec{k}}\right);~~~\vec{\epsilon}_{\vec{k}} = \frac{\vec{k}}{k},
\end{equation}
with explicit forms
\begin{align}\label{eq:sigmaHankel}
        \bar{\sigma}_0 &= \mathcal{A} N_0(-k\eta)^{3/2}H_{i\mu}^{(1)}(-k\eta), \nonumber \\
        \bar{\sigma}_1 &= \frac{i}{2}\mathcal{A}N_0(-k\eta)^{1/2}\left[-H_{i\mu}^{(1)}(-k\eta)-(-k\eta)\left(H^{(1)}_{i\mu+1}(-k\eta)-H^{(1)}_{i\mu-1}(-k\eta)\right)\right],
\end{align}
where $\mu = \sqrt{m^2-1/4}$. The prefactors are determined by normalisation,
\begin{eqnarray}
        &&N_0 = \sqrt{\frac{\pi}{2}}\frac{1}{\sqrt{2k}}\frac{1}{m},\nonumber\\
        &&\mathcal{A} = e^{i\pi/4}e^{-\pi\mu/2}.
\end{eqnarray}

If the particles mass is large compared to Hubble, $\mu\gg H$, we can approximate the mode functions by their adiabatic (WKB) solution
\begin{align}
\label{eq:sigma0adiabatic}
    \bar{\sigma}_0(x=-k\eta) &\approx \frac{1}{m\sqrt{2k}}\frac{x^{3/2}}{(x^2+\mu^2)^{1/4}}e^{i\int^x\sqrt{1+\mu^2/x'^2}dx'},
    \\ \label{eq:sigma1adiabatic}
    \bar{\sigma}_1(x=-k\eta) &\approx -\frac{1}{m\sqrt{2k}}\left(\frac{i}{2}\frac{2x^2+\mu^2}{x^2+\mu^2}+\sqrt{x^2+\mu^2}\right)\frac{x^{1/2}}{(x^2+\mu^2)^{1/4}}e^{i\int^x\sqrt{1+\mu^2/x'^2}dx'}.
\end{align}
We will use these forms instead of \eq{eq:sigmaHankel} henceforth. 
This will greatly simplify the calculation as we will see in \app{App:StationaryPhase}.

%%%%%%%%%%%%%%%%%%%%%%%%%%%%%%%%%%%%%%%%%%%%%%
\section{The full set of interactions}
\label{App:Ints}
The full Lagrangian along with the interaction given in \eq{eq:lagrangian} is
\begin{equation}
    \mathcal{L} = -\frac{1}{2}(\nabla_\mu\phi)^2-V_{\rm s.r}(\phi)-\frac{1}{2}|F_{\mu\nu}|^2 - m^2 |A_\mu^2| + \frac{c}{\Lambda^3}\nabla_\mu\phi A^\mu (\nabla_\rho\phi)^2e^{-i\phi/\Lambda} + h.c. \,,
\end{equation}
where $V_{\rm s.r.}(\phi)$ is the slow roll potential that drives inflation.
It is easiest to calculate the background fields by performing the field redefinition
\begin{equation}
\label{eq:FieldRed}
    A_\mu \mapsto A_\mu + \frac{c^{*}}{m^2\Lambda^3}(\nabla_\rho\phi)^2\nabla_\mu\phi e^{i\phi/\Lambda},
\end{equation}
which removes the interaction term and replaces it with 
\begin{equation}
    \mathcal{L}_{\rm int} = \frac{c}{m^2\Lambda^3}F^{\mu\nu}\nabla_\mu\phi \nabla_\nu((\nabla_\rho\phi)^2)e^{-i\phi/\Lambda} + h.c.
\end{equation}

In this basis, the interaction becomes trivial if all three $\phi$'s are replaced by the background, $\phi_0(\eta)$. To see this, notice that $(\nabla_\rho\phi)^2$ is a scalar, so the second covariant derivative can be replaced by a partial derivative, $\partial_\nu$. If both fields are set to their background configuration, $(\nabla_\rho\phi_0)^2 = (\lambda\Lambda)^2$ is constant. The partial derivative then gives $\mathcal{L}_{\rm int}(\phi_0) = 0$. 
Therefore, after the field definition the vector cannot obtain a VEV. Undoing the field redefinition in \eq{eq:FieldRed} then implies that in the original basis the vector field has a background $\mathcal{A}_{\mu}(\eta)$, which can be found by plugging $\phi_0$ into the shifting term in \eq{eq:FieldRed},
\begin{equation}
    \mathcal{A}_\mu(\eta) = \delta_\mu^0 \frac{c^{*}}{m^2}\frac{\lambda^3}{\eta}(-\eta)^{-i\lambda}.
\end{equation}
Expanding both fields about their background, $\phi \mapsto \phi_0 + \delta\phi$ and $A_\mu \mapsto \mathcal{A}_\mu + A_\mu$, gives the full Lagrangian
\begin{eqnarray}
    \mathcal{L} && = \mathcal{L}_{\rm free} + c(-\eta)\lambda^3(-\eta)^{i\lambda}A_0\left(e^{-i\delta\phi/\Lambda}-1\right) - 2\frac{|c|^2}{m^2}\lambda^6 \cos(\delta\phi/\Lambda)\label{eq:FullLagrangian}\\&&\hspace{0.5cm}\nonumber - \frac{c}{\Lambda}\lambda^2(-\eta)^{i\lambda}\nabla_\mu\delta\phi A^\mu e^{-i\delta\phi/\Lambda} + 2 \frac{c}{\Lambda}\lambda^2\eta^2(-\eta)^{i\lambda}A_0 \delta\phi' e^{-i\delta\phi/\Lambda}\\
    &&\hspace{0.7cm}\nonumber -6 \frac{|c|^2}{m^2}\frac{\lambda^5}{\Lambda}(-\eta)\delta\phi'\cos(\delta\phi/\Lambda) - \frac{c}{\Lambda^2}\lambda(-\eta)(-\eta)^{i\lambda}A_0(\nabla_\rho\delta\phi)^2e^{-i\delta\phi/\Lambda}\\&&\hspace{0.9cm}\nonumber - 2\frac{c}{\Lambda^2}\lambda(-\eta)(-\eta)^{i\lambda}\nabla_\mu\delta\phi A^\mu \delta\phi' e^{-i\delta\phi/\Lambda} + 2\frac{|c|^2}{m^2}\frac{\lambda^4}{\Lambda^2}(\nabla_\rho\delta\phi)^2\cos(\delta\phi/\Lambda)
    \\&&\hspace{1.1cm}\nonumber - 4 \frac{|c|^2}{m^2}\frac{\lambda^4}{\Lambda^2}(\delta\phi')^2 \cos(\delta\phi/\Lambda) + \frac{c}{\Lambda^3}(-\eta)^{i\lambda}\nabla_\mu\delta\phi A^\mu(\nabla_\rho\delta\phi)^2e^{-i\delta\phi/\Lambda}
    \\&&\nonumber\hspace{1.3cm} - 2\frac{|c|^2}{m^2}\frac{\lambda^3}{\Lambda^3}\eta\delta\phi'(\nabla_\rho\delta\phi)^2\cos(\delta\phi/\Lambda) + h.c. \, 
\end{eqnarray}

%%%%%%%%%%%%%%%%%%%%%%%%%%%%%%%%%%%%%%%%%%%%%%
\section{Detailed stationary phase calculation}

\label{App:StationaryPhase}
The leading contribution to the three point function is calculated by expanding \eq{eq:ininmasterformula} to first order in perturbations. As argued in the main text, the leading contribution comes from the $I_{-+}$ diagram
\begin{eqnarray}    I_{-+} = \langle0\rvert \int d^4x_- \sqrt{-g} i \mathcal{H}_{\rm mix}\delta\phi_{\vec{k}_3}\delta\phi_{\vec{k}_1}\delta\phi_{\vec{k}_2}\int d^4x_+ \sqrt{-g}(-i)\mathcal{H}_{A\phi\phi}\lvert 0\rangle + c.c.
\end{eqnarray}
After contracting all fields and factoring out momentum-conserving delta functions, the result factorises into two integrals
\begin{equation}    \label{eq:3point1}I_{-+} = \frac{|c|^2}{\Lambda^3}I_{k_3}^- I_{k_{12}}^+ + c.c.
\end{equation}

The first integral over the two point mixing vertex is
\begin{eqnarray}
\label{eq:Iminus}
    I_{k_3}^- &=& \frac{1}{\sqrt{2k_3^3}}\int\limits_{-\infty}^0 \frac{d\eta}{\eta^4}(-\eta)^{-i\lambda}\left\{(-i\eta\lambda^3\bar{u}_{k_3}+3\eta^2\lambda^2\bar{u}'_{k_3})\bar{\sigma}_0(k_3,\eta)-i\lambda^2\eta^2k_3\bar{u}_{k_3}\bar{\sigma}_1(k_3,\eta)\right\}.
\end{eqnarray}
First insert the mode functions from eqs.~\eqref{eq:scalarmodefunctions}, \eqref{eq:sigma0adiabatic}, and \eqref{eq:sigma1adiabatic}, then make the substitution $x=-k_3\eta$ to massage this into a form appropriate for applying stationary phase
\begin{equation}
    I_{k_3}^- = \frac{\lambda^2}{2\sqrt{2}mk_3^{3/2-i\lambda}}\int_0^\infty dx \frac{C(x)}{x^{3/2}(x^2+\mu^2)^{1/4}}e^{ig(x)}.
    \label{eq:SPA1}
\end{equation}
The two functions are given by
\begin{eqnarray}
    C(x) &=& i(1-ix)\left(\lambda+\sqrt{x^2+\mu^2} + \frac{i}{2}\frac{2x^2+\mu^2}{x^2+\mu^2}\right)-3x^2,\\
    g(x)&=&-\lambda\ln x + x + \int^x \sqrt{1+\mu^2/y^2}dy.
\end{eqnarray}
This integral has a late-time divergence in the limit $x\rightarrow 0$. However, the residual shift symmetry ensures that the late-time divergences cancel in the sum over all Feynman diagrams. 
This is discussed further in \app{App:IRDivergences}.
Since the stationary phase occurs at early time, well away from the $\eta\rightarrow 0$ boundary, the contribution from around the stationary point is still a valid method to evaluate these integrals. 
% More details on this point are given in appendix \ref{App:IRDivergences}. 

The stationary phase approximation $g'(x_-)=0$, yields $x_- = (\lambda^2-\mu^2)/2\lambda$, and $g''(x_-) = 4\lambda^3/(\lambda^4-\mu^4)$ for the time and (inverse of the) variance respectively. The result is 
\begin{eqnarray}
    I_{k_3}^- 
    &\approx& \frac{\lambda^2}{2\sqrt{2}mk_3^{3/2-i\lambda}} \sqrt{\frac{2\pi}{|g''(x_-)|}}\frac{C(x_-)e^{ig(x_-)}}{x_-^{3/2}(x_-^2+\mu^2)^{1/4}},\\
    &=& \frac{\sqrt{\pi}\lambda^2 }{mk_3^{3/2-i\lambda}}\frac{\sqrt{\lambda}}{\lambda^2-\mu^2}C(x_-)e^{i\delta(\mu,\lambda)}
    \label{eq:Ik-}
\end{eqnarray}
where $e^{i\delta(\mu,\lambda)}$ is an overall phase that can be neglected. Notice that the time calculated from stationary phase is identical to that estimated by energy conservation in the locally Minkowski frame, with the mass $m$ replaced by $\mu=\sqrt{m^2-1/4} \approx m$. 

\subsection{Squeezed limit}
The second integral comes from the three-point vertex, and carries the dependence on the shape of the triangle. There is a sum over both contractions of the external legs with the inflaton fields in the vertex
\begin{eqnarray}
    I_{k_{12}}^+ &=& \frac{1}{\sqrt{4k_1^3k_2^3}}\int_{-\infty}^0 \frac{d\eta}{\eta^4}(-\eta)^{i\lambda}\left\{\left(\left(\frac{\lambda^3\eta}{2}+\lambda\eta^3k_1k_2\right)u_{k_1}u_{k_2}-3i\lambda^2\eta^2u_{k_1}u'_{k_2} -3\lambda\eta^3u'_{k_1}u'_{k_2}\right)\right.\sigma_0\nonumber\\&&\hspace{2cm}+\vec{k}_2\cdot \hat{k}_3\left(-\lambda^2\eta^2u_{k_1}u_{k_2}+2i\lambda\eta^3 u'_{k_1}u_{k_2}\right)\sigma_1(k_3,\eta)+(k_1\leftrightarrow k_2) \bigg\}.
    \label{eq:Iplus}
\end{eqnarray}
Equation \eq{eq:Iplus} is solved by again substituting $x=-k_3\eta$, and the mode functions from \eqs{eq:sigma0adiabatic}{eq:sigma1adiabatic}. 

The term multiplying $\sigma_1$ is almost antisymmetric in the exchange of $k_1$ and $k_2$, $\vec{k}_1 \cdot \vec{k}_3 = -k_1k_3\cos\theta$ whereas $\vec{k}_2\cdot \vec{k}_3 = k_2k_3\cos\theta$. In the squeezed limit there is a slight difference between $k_1$ and $k_2$. Using the cosine rule and Taylor expanding
\begin{equation}
    k_2 = k_1\left(1- \frac{k_3}{k_1}\cos\theta + O\left(\frac{k_3}{k_1}\right)^2\right)\implies k_1-k_2=k_3\cos\theta.
\end{equation}

The second integral can be massaged into a similar form to the first
\begin{equation}
    I_{k_{12}}^+ = \frac{\lambda}{2\sqrt{2}mk_1^3k_2^3 k_3^{-3/2+i\lambda}}\int_0^\infty dx \frac{A(x)+B(x)\cos^2\theta}{x^{3/2}(x^2+\mu^2)^{1/4}}e^{ih(x)},
\end{equation}
where 
\begin{eqnarray}
    A(x) &=& -\frac{\lambda^2}{2}-i\lambda^2\frac{px}{2}+\left(\frac{\lambda^2}{2}+3i\lambda-1\right)\left(\frac{px}{2}\right)^2-(6\lambda+2i)\left(\frac{px}{2}\right)^3+4\left(\frac{px}{2}\right)^4\\
    B(x) &=& \frac{1}{2}\left(\sqrt{x^2+\mu^2}+\frac{i}{2}\frac{2x^2+\mu^2}{x^2+\mu^2}\right)\left(\lambda\left(1+\frac{ipx}{2}\right)^2+2i\left(\frac{px}{2}\right)^2\right),\\
    h(x)&=&\lambda\ln x - px - \int^x\sqrt{1+\mu^2/x'^2}dx'.
\end{eqnarray}
where $k_1= k_2$ was taken in all terms besides those with a factor of $k_1-k_2$. 

The stationary phase approximation in the squeezed limit gives $x_+ \approx (\lambda-\mu)/p$, and $|h''(x_+)| \approx p^2/(\lambda-\mu)$. In this integral the phase depends on $p$ since 
\begin{equation}
    h(x_+) = \lambda \ln x_+ - p x_+ - \sqrt{x_+^2+\mu^2} - \mu\ln\left(\frac{x_+}{\mu^2+\mu\sqrt{x_+^2+\mu^2}}\right).
\end{equation}
In the squeezed limit every term is approximately independent of $p$ besides the two logs, $(\lambda-\mu)\ln(x_+)$. The result there has a non-analytic dependence coming from 
\begin{equation}
    e^{ih(x_+)} = p^{i(\mu-\lambda)}e^{i\delta'(\mu,\lambda)}.
\end{equation}

The final result is given by
\begin{eqnarray}
    I_{k_{12}}^+ = \frac{\sqrt{\pi}\lambda e^{i\delta'(\mu,\lambda)}}{2\sqrt{2}mk_1^3k_2^3 k_3^{-3/2+i\lambda}}\frac{A(x_+)+B(x_+)\cos^2\theta}{\sqrt{\mu}(\lambda-\mu)}p^{1/2+i(\mu-\lambda)}.
    \label{eq:Iplusfinal}
\end{eqnarray}
Combining equations \eqref{eq:3point1}, \eqref{eq:Ik-}, and \eqref{eq:Iplusfinal} leads to a result for the three point function
\begin{eqnarray}       
I_{-+}&&\approx \frac{\pi|c^2|}{2\sqrt{2}m^2k_1^3k_2^3}p^{1/2+i(\mu-\lambda)}\frac{\lambda^3}{\Lambda^3}\sqrt{\frac{\lambda}{\mu}}\frac{C(x_-)\left[A(x_+)+B(x_+)\cos^2\theta\right]}{(\lambda-\mu)(\lambda^2-\mu^2)}.
\end{eqnarray}
Equation \eqref{eq:Fsqdef} is then used to relate the three point function to the observable quantity, $F_{\rm sq}$
\begin{equation}
    F_{\rm sq} \approx  f_{\rm oscil}(c,\mu,\lambda)p^{-5/2+i(\mu-\lambda)}+c.c.
       \label{eq:3pointsq}
\end{equation}
where the amplitude of oscillations is given by
\begin{equation}
f_{\rm oscil}(c,\mu,\lambda)= \frac{20\pi|c^2|}{3\sqrt{2}m^2}\dot{\phi}_0\frac{\lambda^3}{\Lambda^3}\sqrt{\frac{\lambda}{\mu}}\frac{C(x_-)\left[A(x_+)+B(x_+)\cos^2\theta\right]}{(\lambda-\mu)(\lambda^2-\mu^2)}.
\label{eq:foscil}
\end{equation}

%\subsection{Power Spectrum}

%To calculate the power spectrum it's useful to notice that the $I_{-+}$ diagram can be related to the square of the two point bispectrum vertex, $I_{k_3}^-$
%\begin{equation}
%    I_{-+} = \frac{|c|^2}{\Lambda^2}|I_{k_3}^-|^2,
%\end{equation}
%The result \eqref{eq:Ik-} is used to obtain an approximate result for the leading correction to the power spectrum
%\begin{equation}  
%\label{eq:PowerSpectrumResult}
%\langle\delta\phi_{\vec{k}}\delta\phi_{-\vec{k}}\rangle' \approx \frac{1}{2k^3}\frac{|c|^2}{\Lambda^2}\frac{4\pi }{m^2}\frac{\lambda^5 |C(x_-)|^2}{(\lambda^2-\mu^2)^2},
%\end{equation}
%where a factor of 2 comes from the complex conjugate, $I_{+-}$ diagram. This expression is used to place a bound on the size of the coupling, $|c|$.
%%%%%%%%%%%%%%%%%%%%%%%%%%%%%%%%%%%%%%%%%%

\section{Cancellation of late-time divergences}
\label{App:IRDivergences}
In this section it is shown that the apparent late-time divergences in calculations due to the non-derivative inflaton couplings are ultimately absent. 
It is demonstrated that the leading late-time divergences cancel in the sum over all Feynman diagrams, and that the stationary phase approximation extracts the non-analytic signal relevant for cosmological collider physics.

\subsection{Justifying the stationary phase approximation}

% Although it appears the inflaton fluctuations have developed a potential, there is still a hidden shift symmetry
% \begin{eqnarray}
% \label{eq:Cshift1}
%     &&\delta\phi \mapsto \delta\phi + \alpha,\\
%     &&A_\mu \mapsto e^{i\alpha/\Lambda}A_\mu + \left(e^{i\alpha/\Lambda}-1\right)\delta_\mu^0 \frac{c^{*}}{m^2}\frac{\lambda^3}{\eta}(-\eta)^{-i\lambda}.
%     \label{eq:Cshift2}
% \end{eqnarray}
% The hidden symmetry ensures that all in-in correlators are IR finite. While individual diagrams may appear to have IR divergences, this symmetry guarantees that they ultimately cancel. 
The divergent integrals appearing in the NG calculation are of the form
\begin{equation}
\label{eq:IRdivint}
    I(p) = p^m\int_0^\infty dx x^{-n \pm i(\mu-\lambda)}e^{ ipx}e^{-if(x,\mu,\lambda)},
\end{equation}
where $x=-k\eta$. 
For $n \geq$ 1, the integrand in \eq{eq:IRdivint} diverges as $x\rightarrow 0$.
In the vector case, $n=3/2$, while in the scalar case of \eq{eq:scalarcoupling}, $n=5/2$ or $3/2$. 
Therefore, it appears that there are late-time divergences in our integrals.
Before we demonstrate finiteness, notice that if we differentiate with respect to the squeezing parameter $p$ enough (say `$k$' times), the effective $n$ can be made small enough such that 
% The integrand in \eq{eq:IRdivint} diverges as $x\rightarrow 0$, but by differentiating with respect to the squeezing parameter $p$ enough number of times, 
the integrand is rendered finite,
\begin{equation}
    \frac{\partial^k}{\partial p^k}I(p) < \infty.
\end{equation}
Therefore, the form of $I(p)$ must be
% By integrating $k$ times
\begin{equation}
    I(p) = I_1 + I_2 p + \cdots + I_k p^{k-1} + \textrm{finite},
\end{equation}
where the $I_i$ are potentially divergent constants of integration. 
From this, we learn that the apparent late-time divergences (which eventually cancel with other diagrams) are analytic in the squeezing parameter, $p$. 
In cosmological collider physics we are interested in the non-analytic dependence, which only comes from the region at early times near the saddle point of \eq{eq:IRdivint}. 
The issue of late-time divergences therefore does not affect the calculation of non-analytic dependence in the diagrams. 
It is therefore justified to only consider the diagrams for which a saddle point exists, or equivalently for which energy conservation can be satisfied at each vertex, and perform the stationary phase approximation, ignoring the region of integration near $x \rightarrow 0$.

The reader may still worry about the validity of the model since the analytic part of the NG seems to diverge. 
Below, we justify why the apparent divergences cancel, at least for the leading non-Boltzmann suppressed contributions.

\subsection{``Integrating out" the vector field}

In the following sections we treat the tadpole of the heavy particle perturbatively.
Here we show that despite the individual diagrams appearing to be divergent, the leading late-time divergences must cancel when all diagrams are summed.
This is a reflection of the shift symmetry
\begin{eqnarray}
\label{eq:Cshift1}
    &&\delta\phi \mapsto \delta\phi + \alpha,\\
    &&A_\mu \mapsto e^{i\alpha/\Lambda}A_\mu,
    \label{eq:Cshift2}
\end{eqnarray}
which protects the slow-roll potential of the inflaton. 

As seen in the main text, the tree level contribution to the inflaton three-point function is of the form
\begin{equation}
    \langle \delta\phi^3\rangle' = \int_{-\infty}^0 \frac{d\eta}{\eta^{5/2}}\int_{-\infty}^0 \frac{d\eta'}{{\eta'}^{5/2}}f(\eta,\eta'),
\end{equation}
where $f$ is some function satisfying $f(0,0)=$ const, and the integrals may or may not be (anti)time-ordered. 
These integrals are divergent at late-times, and one may worry about two possibilities:
\begin{enumerate}
    \item There are regions of integration in which $\eta \gg \eta'$ (or vice versa).
    \item An overall divergence in the integration region $\eta\approx\eta'\rightarrow 0$.
\end{enumerate}
Since we are treating the tadpole perturbatively all diagrams involve a single internal line of the heavy particle. In the first case the particle is propagating for a very long time and therefore must be on-shell. 
The vertex at late times cannot then satisfy energy conservation so these will be exponentially suppressed diagrams.

In the region of integration $\eta\approx\eta'\rightarrow 0$, the particle is propagating for a short time and is therefore off-shell. 
In this scenario we have too much energy ($m < \lambda$) to produce the vector on-shell, and its exchange can be integrated out to give a \textit{local} vertex.
%To understand this we note that in a terrestrial collider we can't integrate out a particle with mass lower than the CoM energy because initial state radiation (ISR) allows the intermediate particle to be placed on-shell.

To understand how this works we go to the local Minkowski frame at $\eta\approx \eta' \rightarrow 0$. We consider the Minkowski space Feynman propagator with 4-momentum, $k$. Due to the $\lambda$ injection we can say that the energy going through the intermediate line is $k_0 = \lambda + \delta k_0$. In this case there is a well-defined expansion of the heavy off-shell propagator
\begin{eqnarray}
    \frac{i}{k^2-m^2} &=& \frac{i}{\lambda^2-m^2 + 2\lambda \delta k_0 + \delta k_0^2 - |\vec{k}|^2}\\
    &=& \frac{i}{\lambda^2-m^2}\left[1 - \frac{2\lambda \delta k_0}{\lambda^2-m^2}+\cdots\right],
\end{eqnarray}
where successive terms are suppressed by $1/(\lambda^2-m^2)$, rather than the usual $1/m^2$. The contributions to a correlator in the region $\eta\approx \eta'\rightarrow 0$ can therefore be reproduced with \textit{local} effective vertices (polynomial in $\delta k_\mu$) involving only $\delta\phi$. The symmetry \eq{eq:Cshift2} is then simply a shift symmetry for $\delta\phi$, leading to manifestly divergence free derivative couplings.

 \bibliographystyle{JHEP}
 \bibliography{references.bib}
% %%%%%%%%%%%%%%%%%%%%%%%%%%%%%%%%%%%%%%%%%%%%%%%%%%%

\end{document}